\documentclass[fleqn,10pt]{wlscirep}

\title{Exact firing time statistics of neurons driven by discrete inhibitory noise}

\author[1,2,3,+]{Simona Olmi}
\author[2,4,5,+,*]{David Angulo-Garcia}
\author[6]{Alberto Imparato}
\author[2,3,4,5,7]{Alessandro Torcini}

\affil[1]{Weierstrass Institute  for Applied Analysis and Stochastics, Mohrenstra$\ss$e 39, 10117 Berlin, Germany}
\affil[2]{Aix Marseille Univ, INSERM, INS, Inst Neurosci Syst, Marseille, France}
\affil[3]{CNR - Consiglio Nazionale delle Ricerche - Istituto dei Sistemi Complessi, 50019 Sesto Fiorentino, Italy}
\affil[4]{Aix Marseille Univ, INSERM and INMED, Marseille, France}
\affil[5]{Aix Marseille Univ, Universit\'{e} de Toulon, CNRS, CPT, UMR 7332, 13288 Marseille, France}
\affil[6]{Aarhus University,Department of Physics and Astronomy, 8000 Aarhus, Denmark}
\affil[7]{Laboratoire de Physique Th\'eorique et Mod\'elisation,
CNRS UMR 8089, Universit\'e de Cergy-Pontoise, F-95300 Cergy-Pontoise Cedex, France}

\affil[*]{david.angulo-garcia@univ-amu.fr}
\affil[+]{these authors contributed equally to this work}

\begin{abstract}
Neurons in the intact brain receive 
a continuous and irregular synaptic bombardment 
from excitatory and inhibitory pre-synaptic neurons,
which determines the firing activity
of the stimulated neuron. In order to investigate the influence of 
inhibitory stimulation on the firing time statistics, we consider Leaky Integrate-and-Fire neurons subject to inhibitory instantaneous post-synaptic potentials. 
In particular, we report exact results for the firing rate, the coefficient of variation and the spike train spectrum for various synaptic weight distributions.
Our results are not limited to stimulations of infinitesimal amplitude,
but they apply as well to finite amplitude post-synaptic potentials, thus
being able to capture the effect of rare and large spikes.
The developed methods are able to reproduce also the average firing properties of heterogeneous neuronal populations. 
\end{abstract}
\begin{document}

\flushbottom
\maketitle
%
%
\thispagestyle{empty}

\section*{Introduction}

Neurons in the neocortex {\it in vivo} are subject to a continuous synaptic bombardment
reflecting the intense network activity~\cite{destexhe2003}.
In the so-called {\it high-input regime}, in which neurons receive hundreds 
of synaptic inputs during each interspike interval~\cite{shadlen1998},
the firing statistics of model neurons is usually obtained in the context of 
the {\it diffusion approximation} (DA)~\cite{ricciardi2013diffusion,tuckwell2005}.

Within such an approximation the post-synaptic potentials (PSPs) are assumed
to have small amplitudes and high arrival rates, therefore 
the synaptic inputs can be treated as a continuous stochastic
process characterized simply by its average and variance, while the shape of the distribution 
of the amplitudes of the PSPs is irrelevant~\cite{roxin2011}.
However several experimental studies have revealed that
rare PSPs of large amplitude can have a fundamental impact in the network activity~\cite{song2005,lefort2009}.
Furthermore, the experimentally measured synaptic weight distributions
display, both for excitatory and inhibitory PSPs, a long tail towards large amplitudes
and a peak at low amplitudes~\cite{miles1990,mason1991,song2005,barbour2007,lefort2009,deweese2006,buzsaki2014}.
The effect of rare and large excitatory post-synaptic potentials (EPSPs) has been
recently examined for generalized leaky integrate-and-fire (LIF) models with generic EPSP
distributions~\cite{iyer2013} and in balanced sparse networks for conductance based LIF neurons
with log-normal EPSP distributions~\cite{teramae2012}. The presence of few strong synapses
induce faster and more reliable responses of the network even for small inputs~\cite{iyer2013,teramae2012}.
Interestingly, in~\cite{teramae2012} it has been shown that a single neuron driven by random
synaptic inputs log-normally distributed reveals a clear aperiodic stochastic resonance~\cite{collins1995,gammaitoni1998},
which is not evident for Gaussian distributed EPSPs.

However, even for the simple case of LIF neurons exact analytic results are still lacking for
large PSPs with generic synaptic weight distributions, apart for the case of the exponentially
distributed PSPs reported in~\cite{richardson2010firing}. In particular, Richardson and Swarbrick have been able to 
obtain the statistics of interspike interval (ISI) for  LIF neurons receiving balanced excitatory and 
inhibitory Poissonian spike trains with exponentally distributed synaptic weights~\cite{richardson2010firing}.
Furthermore, results for generic EPSP distributions have been obtained in ~\cite{iyer2013}
by developing a semi-analytic approach to solve the continuity equation for the membrane
potential distribution.

In this paper, we report exact analytic results for the firing time statistics of neurons receiving inhibitory Poissonian 
spike trains for various synaptic weights distributions. Namely, we estimate the firing time statistics for LIF neurons
subject to inhibitory post-synaptic potentials (IPSPs) (with instananeous rise and decay time)
characterized by constant amplitude, as well as for uniform and truncated Gaussian IPSP distributions. Furthermore, we apply the developed formalism to sparse
inhibitory networks with heterogeneous neuronal properties.

\section*{Models and Methods}

\subsection*{Model and population-based formalism}
\label{sec:Model}

We will consider the firing statistics of a LIF neuron ~\cite{burkitt2006LIFreviewI,burkitt2006LIFreviewII}
subject to a constant external DC current $\mu_0$ and to a synaptic drive $I(t)$, in this case the dynamical evolution of the membrane potential $v$ is given by the following equation
\begin{equation}
\label{eq:ode_orig}
 \frac{dv}{dt}=-\frac{v-\mu_0}{\tau} + I(t) \; .
\end{equation}
where $\tau = 20$ ms is the membrane time constant. The neuron fires
whenever the membrane potential reaches the threshold value $v_{th} = 10$ mV, afterwards the potential is  
reset to the value $v_{re} = 5$ mV. 
The synaptic current $I(t)$ accounts for the linear superposition of the instantaneous 
excitatory and inhibitory PSPs and it can be written as
\begin{equation}
\label{eq:SynapticCurrent}
I(t) = \sum_{\{t_e\}}a_{e}\delta(t-t_{e}) + 
\sum_{\{t_i\}}a_{i}\delta(t-t_{i}) \; ,
\end{equation}
where $a_x$ denote the amplitudes of EPSPs 
 $x=e$ ($a_e > 0$) and IPSPs  $x=i$
($a_i <0$), while the variables $t_{x}$ represent their respective arrival times, which are assumed to be Poissonian distributed with rates $R_x$.  Eqs. 
\eqref{eq:ode_orig} and \eqref{eq:SynapticCurrent} are equivalent to the Stein's model~\cite{stein1965}
with pulse amplitudes randomly drawn from distributions $A_x(a)$. his model, despite its
extreme simplicity, has been shown to be able to provide optimal predictions for the average firing
rates and spike times of cortical neurons~\cite{rauch,jolivet}.

The aim of this paper is to provide exact analytic expressions for the first two moments of the stationary firing 
statistics, namely the average firing rate $r_0$ and the associated coefficient of variation CV, as well as for the
spike-train spectrum (STS) $\hat{C}(\omega)$~\cite{gerstner2002spiking}. To obtain such results we follow the approach 
developed in~\cite{richardson2010firing}, in particular within  a population-based formalism we introduce the probability 
density $P(v,t)$ of the membrane potentials together with the associated flux $J(v,t)$. The continuity equation relating these two
quantities can be written as:
\begin{equation}
\label{eq:continuityEq}
 \frac{\partial P}{\partial t} +\frac{\partial J}{\partial v}= 
 r(t)\left[\delta(v-v_{re}) -\delta(v-v_{th})\right] + \delta(t)\delta(v-v_{re}) \;.
\end{equation}
On the r.h.s. of the above equation are reported the sink (source) term for the neuronal population associated to the
membrane threshold (reset), with $r(t)$ being the instantaneous firing rate
of the population. The last term on the r.h.s. takes into account 
the initial distribution of the membrane potentials, which are assumed to be 
all equal to the reset value at $t=0$~\cite{richardson2008spike}.

The flux $J(v,t)$ can be decomposed in three terms as follows
\begin{equation}
\label{eq:fluxTotal}
 J=-\left(\frac{v-\mu_0}{\tau}\right)P + J_e +J_i \;;
\end{equation}
where the first term on the r.h.s is the 
average drift, while $J_e$ ($J_i$) represents
the excitatory (inhibitory) fluxes originating from the Poissonian  
synaptic drives. The fluxes can be written as 
a convolution of the distribution of the membrane potentials with the synaptic
amplitude distribution, namely
\begin{equation}
\label{eq:flux_integral}
J_x = R_x \int_{-\infty}^{v} dw \enskip P(w,t) \int_{v-w}^{\infty} da \enskip A_x(a) \;.
\end{equation}

The previous set of equations is complemented by the following boundary conditions:
\begin{equation}
J_e(v_{th},t) = r(t) \,, \quad J_{i}(v_{th},t) = 0 \,, \quad P(v_{th},t) = 0 \,;
\end{equation}
and by the requirement that the membrane potential distribution is properly normalized at any time, i.e
$$\int_{-\infty}^{v_{th}} P(v,t) dv = 1 \;.$$

\subsection*{Analytical method to obtain the exact firing time statistics}
\label{sec:laplace}

The estimation of the firing statistics for a LIF neuron subject to shot noise
has proven to be a problem analytically hard to solve 
\cite{tuckwell1976first,giraudo1997jump}. The reason is related to the
overshoots over the threshold $v_{th}$ induced by the finite amplitude 
of the PSPs, which renders difficult the estimation of the membrane potential distribution.
However, it is well known that one of the few cases in which the first passage time problem can be solved,
is represented by exponentially distributed PSP weights, thanks to the memory-less property associated 
with exponential distributions~\cite{kou2003first}. Richardson and Swarbrick \cite{richardson2010firing} 
made use of this unique property to derive the exact solution of the firing rate for the case in which both 
inhibitory and excitatory kick amplitudes are exponentially distributed.
The fact that the only boundary relevant for the first passage time is $v_{th}$, and considering that no trajectory can cross it from above, implies that inhibitory kicks do not contribute to the overshoot and therefore no restriction 
over the distribution of their amplitudes should be in principle imposed in order to obtain an analytic
solution of the problem.

In the following, using the Laplace transform method, we will derive the analytic expressions for 
the firing rates, the coefficient of variation and for the 
spike-train spectrum for various distributions $A_i(a)$ of
the inhibitory amplitudes. For what concerns the excitatory synaptic input,
we will limit our investigation to two analytic solvable cases: namely, 
to exponentially distributed synaptic weights, where
$A_e = \Theta (a) \exp(-a/a_e) / a_e$, and 
to constant excitatory synaptic drive, encompassed in an external 
DC current $\mu_0 > v_{th}$.
 
Let us first consider the excitatory term for exponentially distributed $a_e$, 
in this case the integral equation for the excitatory flux Eq. \eqref{eq:flux_integral}
can be rewritten in a differential form as
\begin{eqnarray}
\label{eq:flux_differential_e}
 \frac{\partial J_e}{\partial v}+ \frac{J_e}{a_e} & = & R_e P(v,t) -r(t)\delta (v-v_{th}) \; ,  
\end{eqnarray}
where 
the last term on the r.h.s. of the above equation accounts for the absorbing boundary condition at threshold.
The bidirectional Laplace transform (from now on only Laplace transform) 
$\tilde{f} (s) = \int_{-\infty}^\infty {\rm e}^{sv} f(v)dv$ of Eq. \eqref{eq:flux_differential_e}
can be written as a combination of linear functions in $\tilde{P}(s,t)$ and $r(t)$, namely
\begin{eqnarray}
\label{eq:J_excit}
\tilde{J}_e(s,t) = \tilde{P}(s,t) Q_e(s) - r(t) S_e(s) \;.
\end{eqnarray}
For the particular case of the exponentially distributed EPSP amplitudes
\begin{eqnarray}
\label{eq:QandS_excit}
Q_e(s) =  \frac{R_e}{1-sa_e}  \qquad S_e (s) =  \frac{{\rm e}^{sv_{th}}}{1-sa_e} \;.
\end{eqnarray}

Whenever the excitatory input is simply given by the DC current $\mu_0$,
we will assume $J_e = Q_e = S_e = 0$ and apply the same formulation that we will expose in
the following. \\

For the distribution of inhibitory amplitudes, the only restriction that we will impose 
is that, one should be able to write the Laplace transform of the inhibitory flux as a linear function of the probability 
density function, namely $\tilde{J}_i(s,t)= \tilde{P}(s,t) Q_i(s)$. 

\subsubsection*{Steady State Firing Rate}
\label{sec:stationary}

Under the above assumptions, we can estimate the Laplace transform of the
sub-threshold voltage distribution $Z_0 \equiv Z_0(s)$, which corresponds to the
generating function for the sub-threshold voltage moments.
Therefore, $Z_0$ can be estimated as the Laplace transform of $P(v,t)$ 
when $v_{th}\to \infty$, implying also that $J = r(t) = 0$. 
In particular, by taking the Laplace transform of Eq.~\eqref{eq:fluxTotal}
together with the assumption that $J = 0$, we obtain
\begin{equation}
\label{eq:dZo_generic}
\frac{dZ_0}{ds} = \tau \left(\frac{\mu_0}{\tau} Z_0 + \tilde{J}_e + \tilde{J}_i \right) \; ;
\end{equation}
where we set $Z_0 = \tilde{P}$.

Since $\tilde{J}_e$ and $\tilde{J}_i$ are linear in $\tilde{P}$, we can rewrite Eq. \eqref{eq:dZo_generic} and solve it as:
\begin{eqnarray}
\label{eq:1/Z0_generic}
\frac{1}{Z_0}\frac{dZ_0}{ds} & =& \tau \left(\frac{\mu_0}{\tau} + Q_e + Q_i \right) \; , \\
\label{eq:Z0_generic}
Z_0 &=& H_e \exp \left(\mu_0 s + \tau \int_0^s ds Q_i \right) \; ,
\end{eqnarray}
where the excitatory contribution is encompassed in the term
\begin{equation}
H_e = \begin{cases} (1-sa_e)^{-\tau R_e} &\mbox{for } A_e(a) = \Theta (a) \exp(-a/a_e) / a_e \\
1 & \mbox{for DC current}  \end{cases} 
\end{equation}

Once we have calculated the generating function, we can solve the stationary case  corresponding to 
$\partial P / \partial t = 0$, 
performing the Laplace transform of the continuity equation \eqref{eq:continuityEq},
which reads as
\begin{equation}
\label{eq:cont_general_stat}
\frac{d\tilde P_0}{ds} = \tilde P_0 \tau \left(\frac{\mu_0}{\tau} + Q_e + Q_i \right) 
- \frac{r_0}{s} \left(e^{sv_{re}} - e^{sv_{th}}  + s S_e \right) \;.
\end{equation}
In our notation, the variables with a zero subscript denote stationary quantities. 
As expected, for $r_0=0$ the function $\tilde P_0$ satisfies Eq. \eqref{eq:1/Z0_generic}, 
therefore we can rewrite the previous equation as
\begin{equation}
\label{eq:cont_general_stat2}
\frac{1}{Z_0}\frac{d \tilde P}{ds}  = \frac{\tilde P}{Z^2_0}\frac{dZ_0}{ds}  
- \tau \frac{r_0}{s Z_0}  F(s)  
   \;.
\end{equation}
where we have indicated with $F(s)$ the function multiplying the term $r_0/s$ in the r.h.s
of Eq.~\eqref{eq:cont_general_stat}. The straightforward solution of Eq.~\eqref{eq:cont_general_stat2} is
\begin{equation}
\label{eq:P/Z_generic} 
\frac{\tilde P(s)}{Z_0(s)} = \tau r_0\int_{s}^{\bar{x}} \frac{dc}{c Z_0(c)} F(c)
 +  \frac{\tilde P(\bar{x})}{Z_0(\bar{x})}\; .
\end{equation}
Notice that, although we have derived an expression for $Z_0$, we have no knowledge of the functional form 
of $\tilde P$. Whenever it is possible to identify an integration limit $\bar{x}$, where the term 
$1/Z_0(\bar{x})$ vanishes, the following exact analytic expression for the stationary firing rate can be obtained
\begin{equation}
\label{eq:firing_rate_generic}
\frac{1}{\tau r_0} = \int_0^{\bar{x}} \frac{dc}{c Z_0(c)}  F(c) \;,
\end{equation}
where we have made use of the normalization condition of the probability densities,
i.e. $\tilde{P}_0(0)=Z_0(0)=1$.\\ 
Our analysis is limited to the two previously reported types of excitatory drive,
because in these two cases an integration limit $\bar{x}$, where
$Z_0^{-1}(\bar{x}) = 0 $,  can be easily found to be 
\begin{equation}
\bar{x} = \begin{cases} 
1/a_e & \mbox{for } H_e = (1-sa_e)^{-\tau R_e} \\
\infty & \mbox{for } H_e = 1  \;.
\end{cases} 
\label{xstar}
\end{equation}
It  should be remarked that in presence of both sources
of excitatory drive, the integration limit can still
be identified whenever $\mu_0 < v_{th}$ and it corresponds to the first one
in \eqref{xstar}, while we have been unable to solve the case when both,
the excitatory drift and the excitatory spike train, can lead the neuron
to fire (see conclusions section for a discussion of this point).

\subsubsection*{First and Second Moment of the First Passage Time Distribution}
\label{sec:time}

Let us now focus on the time dependent evolution of the continuity equation,
in this case, the equation can be solved by performing the Fourier transform in time 
and the Laplace transform in the membrane potential of Eq. \eqref{eq:continuityEq} and \eqref{eq:fluxTotal}.
Namely, we obtain

\begin{equation}
\label{eq:Cont_Fourier_generic}
 \frac{d\hat{P}(s,\omega)}{ds} - \tau \left( \frac{\mu}{\tau} +  Q_e + Q_i
  - \frac{i\omega}{s}\right)\hat{P}(s,\omega) =  -\frac{\hat{\rho}(\omega)}{s}
 F(s)  +  \frac{e^{sv_{re}}}{s} \;;
\end{equation}

where 
$\hat{f} (s,\omega) = \int_{-\infty}^{\infty}  {\rm e}^{-2\pi i \omega t }dt \int_{-\infty}^{\infty}  {\rm e}^{sv} f(v,t)dv$, 
and  $\hat{\rho}(\omega)$ is the Fourier transform of the spike-triggered rate. 
Dividing both sides by $Z_0$ and integrating the functions over the interval $s = [ 0, \bar{x}]$, the l.h.s of Eq. 
\eqref{eq:Cont_Fourier_generic} vanishes and an analytic expression for $\hat{\rho}(\omega)$ can be obtained, namely
\begin{equation}
\label{eq:rho_generic}
 \hat{\rho}(\omega) = 
 \frac{\displaystyle \int_{0}^{\bar{x}} ds s^{i\omega\tau} \frac{d}{ds}A(s)}
 {\displaystyle \int_{0}^{\bar{x}} ds s^{i\omega\tau}\frac{d}{ds}B(s)} \;;
\end{equation}
where
\begin{equation}
\label{eq:As_Bs} 
A(s)  =   \frac{e^{sv_{re}}}{Z_0} \,; \qquad B(s)  =  \frac{F(s)}{Z_0} \;.
\end{equation}
Equation \eqref{eq:rho_generic} diverges exactly at $\omega \equiv 0$, and 
in that case it should be complemented with the expression $\hat \rho(\omega = 0) = r_0 \pi \delta(\omega)$. 
The spike-triggered rate (also called the conditional firing rate) provides all the information on the spike train statistics. For instance, 
the spike train spectrum $\hat{C}(\omega)$, which is the Fourier transform of the auto-correlation function of the spike train, is related to the 
spike triggered rate via the formula $\hat{C}(\omega) = r_0[1 + 2 \text{Re}(\hat{\rho}(\omega))]$~\cite{gerstner2002spiking}.\\

Moreover, $\hat{C}(\omega)$ and $\hat{\rho}(\omega)$ are also related with the Fourier transform of the first-passage time density 
$\hat{q}(\omega)$ as follows:
\begin{equation}
\hat{q}(\omega) = \frac{\hat{\rho}(\omega)}{1 + \hat{\rho}(\omega)} \; ; \quad
\hat{C}(\omega) = r_0 \frac{1-|\hat{q}(\omega)|^2}{(1-\hat{q}(\omega))^2} + 2 \pi r_0^2 \delta(\omega) \; ,
\label{eq:Cw}
\end{equation}
where $\lim_{\omega \to 0} \hat{C}(\omega) = r_0$ and $\lim_{\omega \to \infty} \hat{C}(\omega) = \text{CV}^2 r_0$ ~\cite{lindner2002}.

Therefore the $n-th$ moment of the  first passage time distribution  is given by
\begin{equation}
\label{eq:fOmega}
\frac{\partial^n \hat{q}}{\partial \omega^n} |_{\omega = 0} 
= (-\mathrm{i})^n \langle t^n \rangle \; .
\end{equation}
For the estimation of the first two moments it is sufficient to expand to the second order
in $\omega$ the terms entering in Eq.~\eqref{eq:rho_generic}, in particular 
$s^{i \omega \tau} \approx 1 + i \tau \omega \log(s) - \frac{1}{2} (\omega \tau \log s)^2 + \mathcal{O}(\omega^3)$. 
Finally, $\hat{\rho}(\omega)$ can be approximated as a polynomial in $\omega$, that reads as
\begin{equation}
\hat{\rho}(\omega) \approx \frac{n_0 + n_1 \omega + n_2 \omega^2}{d_0 + d_1 \omega + d_2 \omega ^2} \quad;
\end{equation}
where $n_0 = -1$, $d_1 = i/r_0$, $d_0 = 0$ and
\begin{equation}
\label{eq:ns_ds}
n_1  =  i \int_0^{\bar{x}} ds \log s \frac{d}{ds}A(s) \quad d_2  =  \int_0^{\bar{x}} ds \frac{\log s}{s}B(s) \;.
\end{equation}

It is worth to notice that, despite the expansion is limited to the second order, the solutions for the first and second moments
are exact since terms of higher order disappear when evaluated at $\omega = 0$. \\

From the expression \eqref{eq:fOmega} it is easy to verify that 
the first and second moments of the first passage time distribution are given by
\begin{equation}
\label{eq:moments}
\langle t \rangle = -i d_1 = 1/r_0  \quad
\langle t^2 \rangle = 2 \frac{d_1^2 - d_2 n_0 + d_1 n_1}{n_0 ^2} \quad ;
\end{equation}
and from these it is straightforward to estimate the coefficient of variation CV 
of the ISI, namely
\begin{equation}
\label{eq:cv}
\text{CV} = \frac{\sqrt{\langle t \rangle^2 - \langle t^2 \rangle}}{\langle t \rangle} \,.
\end{equation}

As can be seen by this general solution, the two relevant quantities that define uniquely the stationary firing statistics are the functions $Z_0(s)$ and $F(s)$ defined in the Laplace space, where $F(s)$ depends only on the considered
excitatory drive, while $Z_0(s)$ on the whole sub-threshold input.

\subsubsection*{Explicit expressions of $Z_0$ for selected distributions}

In this manuscript, we focus on the exact solutions for four relevant 
types of distributions of the inhibitory synaptic weights $A_i(a)$: namely,
exponential (ED), $\delta$ (DD), uniform (UD) and truncated Gaussian distribution (TGD).
The exact expressions for $Z_0(s)$ for each of these four cases are reported in the following,
together with the average and variance of the corresponding distributions $A_i(a)$.\\


\textbf{\textit{Exponential Distribution (ED):}} In this case the synaptic weight distribution is of the form:

\begin{equation}
A_i(a) = -\frac{\Theta (-a)}{{a}_i} {\rm e}^{-a/{a}_i} \;.
\end{equation}

For this distribution, the equations for the inhibitory flux (Eq. \eqref{eq:flux_integral}) can be rewritten in a differential form analogous to Eq. \eqref{eq:flux_differential_e}, namely 
\begin{eqnarray}
 \label{eq:flux_differential_i}
\frac{\partial J_i}{\partial v}+ \frac{J_i}{a_i} = R_i P(v,t)  \;.
\end{eqnarray}
where the term accounting for the absorbing boundary is not present since we are 
considering the inhibitory shot noise.

The Laplace transform of this equation reads as
\begin{equation}
\label{eq:inhib_flux_expon_laplace}
\tilde{J}_i  =  \frac{R_i}{1-s a_i} \tilde{P} \;.
\end{equation}

For this choice of distribution, it is possible to obtain an explicit expression for
the generating function  the sub-threshold voltage moments, namely
\begin{eqnarray}
Z_0(s) &=& H_e\frac{{\rm e}^{\mu_0 s}}{(1-sa_i)^{R_i \tau}} \;.
\end{eqnarray}

The mean and variance due to the distribution of the inhibitory synaptic
weights for the exponential case are $\langle a_i \rangle = a_i$ and $var[a_i] = a_i^2$;
while the absolute value of the skewness is two.\\


\textbf{\textit{$\delta$ Distribution (DD):}} 
In the case in which the inhibitory population delivers spikes with 
a constant amplitude $a_i$, the distribution becomes simply a $\delta$-function: 
\begin{equation}
A_i(a) = \delta (a-a_i) \;.
\end{equation}

In this case the equation for the inhibitory flux becomes the following
\begin{equation}
\label{eq:FluxInh_delta}
\frac{\partial J_i}{\partial v}=R_i \left[ P(v,t)- P(v-a_i,t)\right]. 
\end{equation}
and the associate Laplace transform reads as 
\begin{equation}
\tilde{J}_i  =  - R_i \tilde{P} \frac{(1-{\rm e}^{s a_i})}{s} \;.
\end{equation}

For this simple distribution the generating function $Z_0$ reads as:
\begin{equation}
Z_0(s) = H_e \frac{\exp(\mu_0 s + \tau R_i E_I(a_i s) )}{C_0 s^{\tau R_i}} \; ;
\end{equation}
where $E_I(y) = \int_{y}^{\infty}dy e^{y}/y$ is the exponential integral 
and $C_0$ accounts for the normalization condition requiring that $Z_0(0) = 1$
and its explicit expression is
\begin{equation}
C_0 = \exp(\tau R_i (\Gamma + \log |a_i|)) \; ,
\end{equation}
where $\Gamma \approx 0.577731$ is the Euler-Mascheroni constant. 

For this distribution the mean and variance of the inhibitory process are simply
$\langle a_i \rangle = a_i$ and $var[a_i] = 0$, and also the skewness is zero. \\


\textbf{\textit{Uniform Distribution (UD):}} 
For uniformly distributed synaptic weights with support $[l_1 \; l_2]$, we can 
express the UD as follows  

\begin{equation}
A_i(a) = \frac{\Theta(a-l_1) - \Theta(a-l_2)}{l_2 - l_1} \;.
\end{equation}

The variation of the flux respect to the potential is given by:
\begin{equation}
\label{eq:FluxInh_Unif}
\frac{\partial J_i}{\partial v}= \frac{R_i}{l_2-l_1} \int_{-\infty}^{v} dw P(w,t)\left[(v-w-l_2)\Theta(v-w-l_2) - (v-w-l_1)\Theta(v-w-l_1) \right]
\end{equation}
and the corresponding Laplace transform reads as
\begin{eqnarray}
\tilde{J}_i & = &  -\tilde{P} \frac{R_i}{l_2-l_1} \left[ \frac{{\rm e}^{l_1s}}{s^2} - \frac{{\rm e}^{l_2s}}{s^2} + \frac{l_2 - l_1}{s} \right]\;.
\end{eqnarray}

This leads to the generating function together with the normalization constant:
\begin{eqnarray}
Z_0(s) & = & H_e \frac{\exp\left( \mu_0 s + \frac{\tau R_i}{l_2 - l_1} 
\left(l_2 E_I(l_2 s) - l_1 E_I(l_1 s) + \frac{{\rm e}^{l_1 s}}{s} - \frac{{\rm e}^{l_2 s}}{s}\right) \right)}{C_0 s^{\tau R_i}} \\
C_0 & = & \exp\left(\frac{\tau R_i}{l_2 - l_1}(l_1 (1 - \Gamma - \log|l_1|) - l_2 (1 - \Gamma - \log|l_2| ))\right) \;
\end{eqnarray}

The mean and variance of the inhibitory shot noise 
are $\langle a_i \rangle = (l_2 + l_1)/2$ and $var[a_i] = (l_2 - l_1)^2/12$, respectively; 
while the skewness is zero.\\


\textbf{\textit{Truncated Gaussian Distribution (TGD):}} 
A biologically relevant distribution is the Gaussian distribution~\cite{isope2002},
which is peaked at $a_p$ and with a standard deviation equal to $\sigma_G$. For notation simplicity we write the Gaussian
distribution as

\begin{equation}
\phi(y) = \frac{1}{\sqrt{2\pi}\sigma_G}\exp\left(-\frac{(y-a_p)^2}{2 \sigma_G^2}\right) \; .
\end{equation}
Since we are only interested in the inhibitory kicks, we truncate the original distribution and we impose 
the support $(-\infty , 0]$. The distribution of the synaptic weights can be written as

\begin{equation}
A_i(a) = \frac{\Theta(-a)\phi(a)}{\Phi(0)} \; ;
\end{equation}

$\Phi(0)$ is the cumulative distribution of the normal distribution evaluated at the upper limit of the 
support according to the equation
\begin{equation}
\Phi(y) = \frac{1}{2}\left(1 + \text{erf} \left(\frac{y-a_p}{\sigma_G \sqrt{2}}\right)\right) \;.
\end{equation}

The flux of inhibitory probability takes the form

\begin{equation}
\label{eq:FluxInh_Gauss}
\frac{\partial J_i}{\partial v}=\frac{R_i}{\Phi(0)}  \left(\Phi(0) P(v,t) 
  + \int^{v}_{-\infty} dw P(w,t) \phi(v-w) \Theta(v-w)\right). 
\end{equation}

The corresponding Laplace transform is:
\begin{equation}
\label{eq:Laplace_Gauss}
\tilde{J}_i = - \tilde P \frac{R_i}{\Phi(0)s}\left(\Phi(0)- \frac{1}{2}\text{erfc}\left( \frac{s\sigma_G^2+a_p}{\sqrt{2}\sigma_G} \right){\rm e}^{s(s\sigma_G^2 + 2a_p)/2}\right)
\end{equation}
The complicated expression appearing in Eq. \eqref{eq:Laplace_Gauss} does not allow us to obtain the
explicit expression for $Z_0$. However it is easy to integrate numerically Eq. \eqref{eq:Z0_generic} once we 
know $\tilde{J}_i$. When dealing with a width of the Gaussian distribution $\sigma_G > 1$, the exponential term 
in Eq. \eqref{eq:Laplace_Gauss} can grow very rapidly while the complementary error function tends to 0, generating 
numerical problems due to machine precision, specially in the evaluation of the terms $s>1$. In such cases one can use 
the first order expansion of the complementary error function: $\text{erfc}(y) \approx \exp(-y^2)/\sqrt{\pi} y$ 
when $y>>1$ and Eq. \eqref{eq:Laplace_Gauss} can be simplified in such cases to:

\begin{equation}
\label{eq:Laplace_Gauss_simplified}
\tilde{J}_i \approx - \tilde P \frac{R_i}{\Phi(0)s}\left(\Phi(0) + \sigma_G \frac{{\rm e}^{-\frac{a_p^2}{2\sigma_G^2}}}{\sqrt{2 \pi}\sigma_G^2 s + a_p} \right) 
\quad \text{if} \; s,\sigma_G > 1
\end{equation}

For the TGD, the mean value and variance of the membrane potentials 
associated to the inhibitory shot noise take the following form
\begin{eqnarray}
\langle a_i \rangle &=& a_p -  \sigma^2_G \frac{\phi(0)}{\Phi(0)} \\
var[a] &=& \sigma_G^2 \left(1 + a_p \frac{\phi(0)}{\Phi(0)} + 
\left( \sigma_G \frac{\phi(0)}{\Phi(0)}\right)^2 \right)\; ;
\end{eqnarray}

The value of the negative skewness grows with $\sigma_G$, namely it passes from $\simeq -0.22$ for $\sigma_G = |a_i/2|$ to $\simeq -0.92$ for $\sigma_G = |5 a_i|$ and it tends to the value $-\sqrt{2}(4-\pi)/(\pi-2)^{3/2} \simeq -0.9992$ for 
$\sigma_G \to \infty$.

\subsection*{Effective Input and  Synaptic Noise Intensity}
\label{sec:NoiseIntensity}

In order to perform meaningful comparisons between the neuronal
response for different synaptic weights distributions, we will consider
the responses obtained for the same \textit{effective}  average 
input $\mu_T$ and noise intensity $\sigma$.

For a neuron receiving an inhibitory Poissonian spike train 
at a rate $R_i$ and with an average synaptic weight $\langle a_i \rangle$ the
effective average input is
\begin{equation}
\label{eq:average_Input}
\mu_T = \mu_e + \mu_i = \mu_e + \tau R_i \langle a_i \rangle \;.
\end{equation}
where $a_i < 0$ and $\mu_e$ is the average excitatory input. When only the drift term is present $\mu_e = \mu_0$, 
while for a neuron receiving also an excitatory Possonian spike train 
of rate $R_e$ and with average synaptic weight $\langle a_e \rangle$, it becomes 
$\mu_e = \mu_0 + \tau R_e \langle a_e \rangle$.
On the one hand, for an effective sub-threshold input $\mu_T < v_{th}$ the dynamics
of the LIF neuron is characterized by two timescales: the relaxation time from
the reset value $v_{re}$ to the resting value $\mu_T$ and the activation
time associated to the escape process from the resting state to the threshold
induced by the fluctuations in the input~\cite{lindner2002}. On the other hand,
for a supra-threshold LIF neuron for which $\mu_T > v_{th}$, in absence of a refractory state, 
the only characteristic time is the tonic firing rate 
\begin{equation}
\label{eq:LIF_simple}
\frac{1}{r_0^t} = \tau \ln \left(\frac{\mu_T - v_{re}}{\mu_T - v_{th}} \right) \;.
\end{equation}

Moreover, in the set-up that we are studying, the neuron is in general subjected to two sources of randomness.
A first source due to the variability of the arrival times of the Poisson process, and a second source due to 
the distribution of the amplitudes of the synaptic weights. Therefore, the total noise intensity $\sigma^2$ 
associated to these two uncorrelated processes is given by the sum of the variances of each process, namely 
\begin{equation}
\label{eq:noise_intensity}	
\sigma^2 = \sum_{x=i,e} R_x \tau (\langle a_x \rangle^2 + var[a_x])
\enskip .
\end{equation}

From Eqs. \eqref{eq:average_Input} and \eqref{eq:noise_intensity} one can see that the values of
$\mu_T$ and $\sigma$ can be independently tuned with an appropriate selection of the parameter $\mu_e$ and
$R_i$ for any fixed average inhibitory amplitude. Whenever the excitatory drive is given simply by a DC term
$\mu_0$, for any choice of the couple of values \{$\mu_T, \; \sigma\}$ one has an unique solution, which
however does not always correspond to a firing activity. For instance, at small values 
of $\sigma^2$ one can find values of $\mu_0<v_{th}$, meaning that the neuron will never fire. This implies
the existence of a firing onset threshold for this class of systems which depends on the shape of the distribution.
Conversely, when exponentially distributed excitatory kicks are present, two parameters 
(amplitude and the rate of arrival of the excitatory kicks) define $\mu_e$ and therefore a whole family of 
solutions exist for any fixed $a_i$. Therefore in this case there is not a finite firing onset threshold,
allowing for arbitrarily small rate even for small noise intensities.\\

Throughout this paper we will compare the results obtained within 
the shot noise framework against the widely used diffusion 
approximation~\cite{ricciardi2013diffusion}. Within such approximation 
the particular shape of the synaptic weight distributions 
are irrelevant and indeed the only relevant quantities defining 
the stationary firing rate and the CV are $\mu_T$ and $\sigma^2$ through 
the formulas
\begin{eqnarray}
\label{eq:r0_diffusion}\frac{1}{r_0} &= & \tau \sqrt{\pi}\int^{y_\theta}_{y_r} dx {\rm e}^{x^2}(1+\text{erf}(x)) \\
\label{eq:CV_diffusion}\text{CV}^2 &=& \frac{2 \pi \tau^2}{r_0^2} \int^{y_\theta}_{y_r} dx {\rm e}^{x^2} \int_{-\infty}^{x} dy {\rm e}^{y^2}(1+\text{erf} (y))^2
\end{eqnarray}
where $y_{\theta}=(v_{th}-\mu_T)/\sigma$ and $y_{r}=(v_{re}-\mu_T)/\sigma$.\\

By following \cite{richardson2010firing}, Eqs. \eqref{eq:r0_diffusion} and \eqref{eq:CV_diffusion} can be obtained 
from the shot noise formulation, namely from Eq.~\eqref{eq:moments} and Eq.~\eqref{eq:Cw}
by considering the expansion of $Z_0$ limited to the first two voltage moments:
\begin{equation}
Z_0(s) \simeq \exp\left(s \mu_T + \frac{s^2}{4} \sigma^2\right) \quad .
\label{z0exp} 
\end{equation} 
 
\section*{Results}

In this Section we will apply the developed formalism to estimate the response of a single LIF neuron
as well as the firing characteristics of a sparse inhibitory neural network for different synaptic
weight distributions. In particular, to verify the limits of applicability of the approach
we will compare the theoretical estimations with numerical data and with the DA.

Usually, the firing time statistics of a neuron subject to a noisy uncorrelated input is theoretically estimated
within the so-called DA~\cite{ricciardi2013diffusion,BrunelHakim1999,tuckwell2005}. 
This approximation is only valid, however, when the PSP amplitudes are small compared with the reset-threshold voltage 
distance and the arrival frequencies are sufficiently high. Outside of such limits the  DA fails to reproduce the numerical data and in particular it is unable to capture the differences due to different synaptic
weight distributions~\cite{richardson2010firing,iyer2013}. Furthermore, the DA 
has been employed to reproduce network activity of recurrent networks, in such a case one should assume
that the spike trains impinging on the neuron are temporally uncorrelated, a condition usually fulfilled 
in sparse networks~\cite{amit1997,BrunelHakim1999}. However this should be considered as a first approximation,
indeed correlations are present even in sparse balanced networks and they can be captured by 
driving a single neuron with a colored Gaussian noise self-consistently generated~\cite{dummer2014}.

\subsection*{Influence of the IPSP distributions on the firing statistics}

\begin{figure}
\centering
\includegraphics[width=1\linewidth]{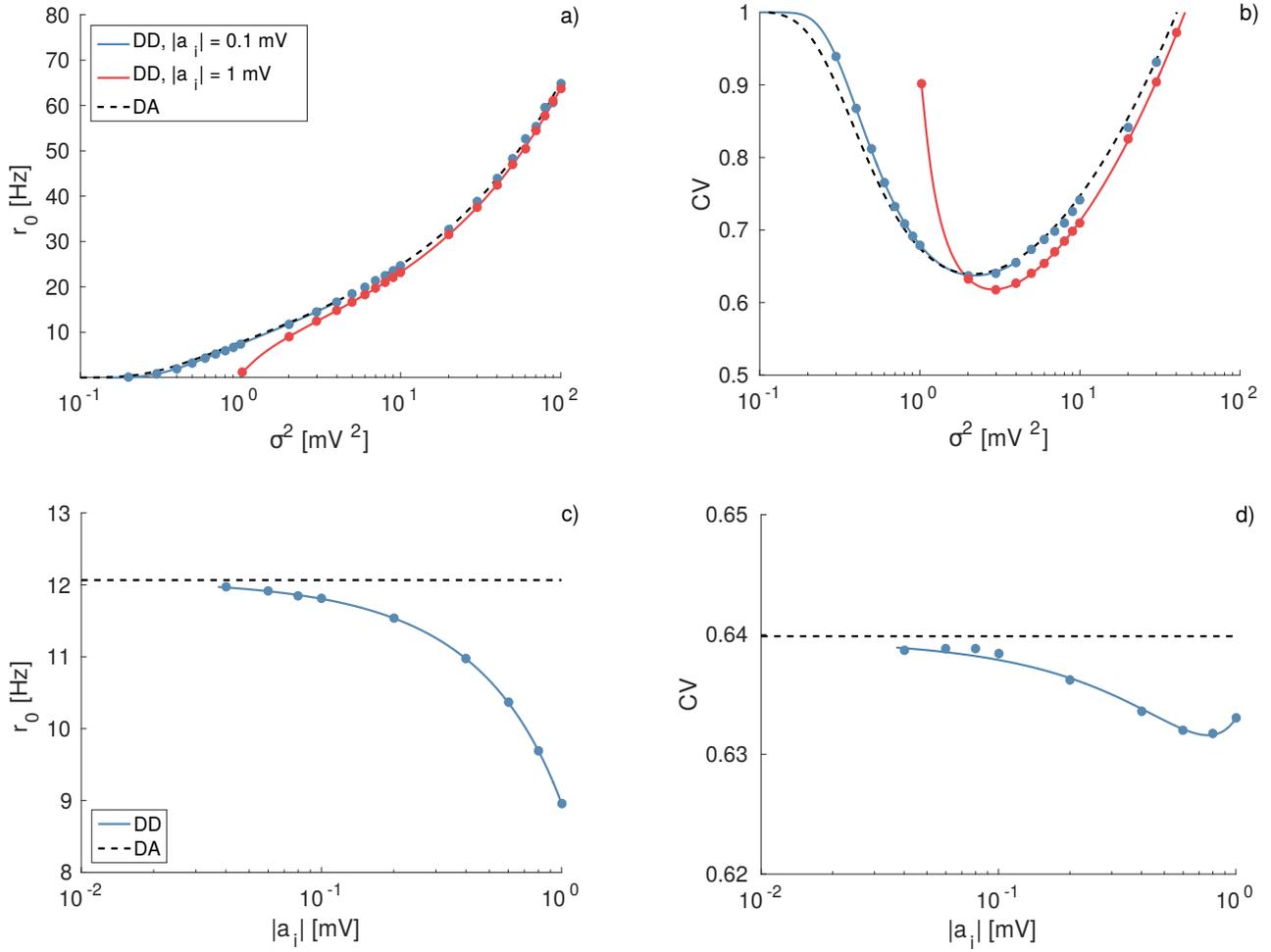}
\caption{{\bf Firing time statistics for different IPSP amplitudes with $\delta$-distributions (DD).} Average firing rate $r_0$ (a) 
and coefficient of variation CV (b) as a function of noise intensity $\sigma^2$. The red curves/symbols correspond 
to $|a_i| = 1$ mV and the blue ones to $|a_i| = 0.1$ mV. Firing rate $r_0$ (c) and CV (d) as a function of the 
synaptic amplitude $|a_i|$ for constant noise intensity, namely $\sigma^2 = 2$ mV$^2$. Black dashed curves refer to 
the diffusion approximation (DA) (Eqs. \eqref{eq:r0_diffusion} and \eqref{eq:CV_diffusion}); solid  curves  
to the theoretical results for shot noise with constant amplitude $a_i$  (Eqs. \eqref{eq:firing_rate_generic} 
and \eqref{eq:cv}) and symbols to the corresponding numerical simulations. Simulations were performed by exactly 
integrating Eq. \eqref{eq:ode_orig} with an event driven 
scheme (see \cite{olmi2016} for details) and the statistics were estimated over $\approx 10^7$ spikes. For all the panels the effective input is $\mu_T = 9.0$ mV and the excitatory drive is
simply a DC term; other parameter values are $v_{re} = 5$ mV, $v_{th}=10$ mV, $\tau = 20$ ms. 
}
\label{fig:Diffusion}
\end{figure}

As a first aspect, let us consider the dynamics of a single LIF neuron subject to an excitatory DC current 
$\mu_e = \mu_0 > v_{th}$ plus the inhibitory contribution given by a Poissonian train of IPSPS with constant amplitude $a_i$. 
In particular, we examine the neuronal response by increasing the noise intensity $\sigma^2$ at a constant effective input current, 
namely $\mu_T = 9$ mV, corresponding to a sub-threshold case. 
In particular, the noise intensity is varied by adjusting the rate of the inhibitory train $R_i$. The results, 
reported in Fig.~\ref{fig:Diffusion}  (a), show that the neuron fires only for sufficiently large noise and the firing rate
$r_0$ increases with $\sigma^2$ as expected. Furthermore, as shown in Fig.~\ref{fig:Diffusion}  (b) the coefficient of variation
exhibits a clear minimum at an intermediate noise amplitude $\sigma_m^2$, 
an effect known as coherence resonance (CR) and widely studied in the context of excitable systems~\cite{lindner2004}. 
The emergence of CR is related to the presence of 
at least two competing time scales which depend differently on the noise amplitude,
for the sub-threshold LIF neuron these two time scales are the relaxation and escape times~\cite{lindner2002}.

For sufficiently small IPSP amplitudes the agreement between the DA, the numerical results and the analytic expression 
reported in Eq.~\eqref{eq:firing_rate_generic} is almost perfect
as evident from Fig.~\ref{fig:Diffusion} (a-b), where the blue symbols/curve 
refer to $|a_i|=0.1$ mV. However, by increasing the IPSP amplitude the agreement between
numerical results and the estimation given by the shot-noise result~\eqref{eq:firing_rate_generic} 
remains very good, while the DA is unable to capture the effect of large IPSP.
In particular, for $|a_i| = 1$ mV the DA fails 
in reproducing the onset of the firing activity as well as the 
position and height of the minimum of the CV, and in general the 
firing statistics for low noise amplitudes
(see red curves/symbols in Fig.~\ref{fig:Diffusion} (a-b)).
Unitary IPSPs of amplitude $\simeq 1$ mV have been measured
experimentally in the hippocampus of Guinea-pig {\it in vitro}~\cite{miles1990}.

As  shown in Fig.~\ref{fig:Diffusion} (a-b), the increase of the IPSP amplitude has a noticeable
effect on the neuronal response, even if 
the effective input $\mu_T$ and the noise intensity
are the same as in the case $|a_i| = 0.1$ mV.  Increasing IPSP amplitudes
leads to a decrease in the firing activity and it induces an increase of the maximal coherence
observable at $\sigma_m^2$, which now occurs at larger noise amplitude
with respect to the case $|a_i| = 0.1$ mV. This can be explained by the fact
that the relaxation times to the equilibrium value $\mu_T$ are longer for
larger IPSP and this induces a deeper minimum in the CV as reported in~\cite{pikovsky1997}.
Furthermore the irregularity in the emitted spikes, as measured
by the CV, increases for $\sigma^2 < \sigma_m^2$  and becomes
more regular at larger noise intensities.

The effect of the IPSP amplitude can be better appreciated by performing
a different test, namely by maintaining constant both the noise intensity 
and $\mu_T$ while increasing $|a_i|$. 
These two quantities can be independently 
tuned with a suitable selection of $R_i$ and $\mu_0$ in Eqs. \eqref{eq:average_Input} 
and \eqref{eq:noise_intensity}.
The results of this analysis are reported
in Fig.~\ref{fig:Diffusion} (c-d), the firing rate exhibits a dramatic
decrease for increasing $|a_i|$, as expected due to the increase of
average inhibitory current $\mu_i$. This effect is
well reproduced by Eq.~\eqref{eq:firing_rate_generic}, but it
is absolutely not captured by the DA as
shown in Fig.~\ref{fig:Diffusion} (c). The increase
of the IPSP amplitude leads to a small variation of the CV
revealing a minimum at an intermediate value $|a_i| \simeq 0.9$ mV
(see Fig.~\ref{fig:Diffusion} (d)). Nevertheless,
the DA provides a constant value for the CV in the
whole examined range. 
The origin of the minimum can be understood observing
Fig.~\ref{fig:Diffusion} (b): by increasing $|a_i|$ the overall minimum
of the CV curve shifts towards larger noise amplitudes, and at the same time
the CV values decrease (increase) for  $\sigma^2 > \sigma_m^2$
($\sigma^2 < \sigma_m^2$). In Fig.~\ref{fig:Diffusion} (d), we consider
a noise intensity $\sigma^2 = 2$ mV$^2$ for all the simulations.
For small (large) $|a_i|$ the maximal coherence is observable at 
$\sigma_m^2 < 2$ mV$^2$ ($\sigma_m^2 > 2$ mV$^2$), thus
the CV value at $\sigma^2 = 2$ mV$^2$ decreases (increases) with $|a_i|$.
The minimum in Fig.~\ref{fig:Diffusion} (d) occurs exactly
when the CV displays its absolute minimum  at $\sigma_m^2 = 2$ mV$^2$.

\begin{figure}
\centering
\includegraphics[width=1\linewidth]{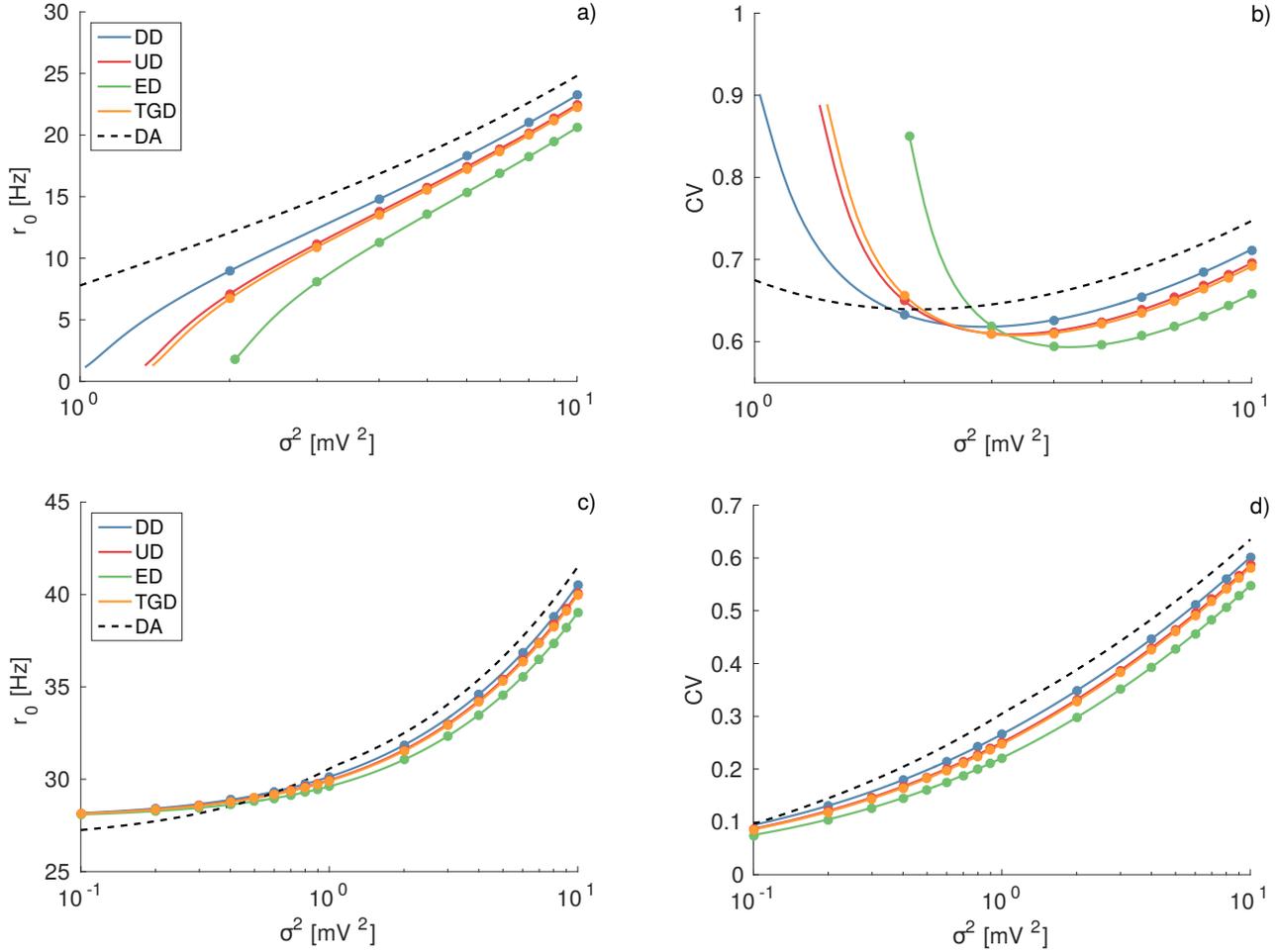}
\caption{{\bf Firing time statistics for different IPSP distributions.}
Firing rate $r_0$ and CV  as a function of noise intensity for sub-threshold 
($\mu_T = 9$ mV) (a,b) and supra-threshold  neurons ($\mu_T =11$ mV) (c,d).
In all the panels the symbols correspond to numerical simulations, the dashed black lines to the DA, calculated from Eqs. \eqref{eq:r0_diffusion} and \eqref{eq:CV_diffusion}, and the solid lines 
to the theoretical results reported in Eq.~\eqref{eq:firing_rate_generic}
and Eq.~\eqref{eq:cv}. The average IPSP amplitude is set to $\langle|a_i|\rangle = 1$ mV for all the distributions.
For the TGD: the peak position $|a_p|$ and the width $\sigma_G$ of the 
distribution are equal (namely, $\approx 0.7766$ mV). For the UD: $l_1 = 2 a_i$
and $l_2 = 0$. Other parameters and simulation procedures as in Fig. \ref{fig:Diffusion}. }
\label{fig:Supra_All}
\end{figure}

For the moment we have considered only the case of constant IPSP amplitudes,
now we will examine the influence of different distributions $A_i$ on the
response of the single LIF neuron. In general, we observe that
the shape of the IPSP distribution can noticeably influence the firing rate and the CV.
In order to verify this observation, we consider the firing statistics 
of a neuron subject to the same average input and noise intensity
obtained by considering inhibitory spike trains with
IPSP distributions of different shapes, but with the same average amplitude $\langle|a_i|\rangle$.
In particular, more asymmetric distributions,
characterized by higher skewness and presenting longer tails towards larger IPSP amplitudes, induce lower 
firing rates, as shown in Fig.~\ref{fig:Supra_All} (a) and (c)
for supra-threshold and sub-threshold neurons, respectively.
In the sub-threshold 
case this implies that passing from $\delta$-distributions, to uniform, to truncated Gaussian and to exponential ones,
the firing onset will occur for larger and larger $\sigma^2$.
For sub-threshold neurons the finite IPSP amplitude enhance the coherence resonance effect with respect
to infinitesimal IPSP (corresponding to the DA), 
while the long inhibitory tails induce a shift of the minimum in the CV
towards larger noise amplitudes (see Fig.~\ref{fig:Supra_All} (b)).
For supra-threshold neurons the increased asymmetry in the distributions simply
induces more regular firing, as shown in Fig.~\ref{fig:Supra_All} (d).
It is quite peculiar that the TGD and the UD give almost identical results,
despite the fact  that the TGD is more asymmetric and characterized by a larger
standard deviation, namely 0.61 mV for the TGD and 0.29 mV for the UD.
Differences are instead seen with respect to the ED which reveals extremely long tails
and a standard deviation of 1 mV  and to the DD where there is no variability in the IPSP
amplitude.  

\begin{figure}
\centering
\includegraphics[width=1\linewidth]{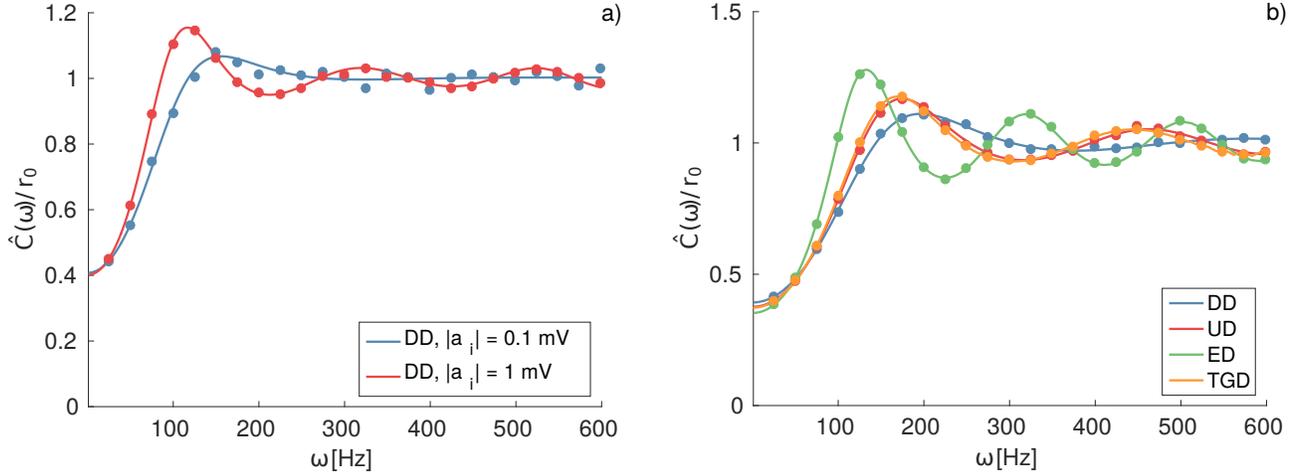}
\caption{{\bf Spike Train Spectra} a) Normalized spike train spectra 
${\hat C}({\omega})/r_0$ for the same cases 
shown in Fig. \ref{fig:Diffusion} (a,b) for an effective input ${\mu_T} = 9$ mV
and noise intensity $\sigma^2 = 2$ mV$^2$;
b) Normalized spike train spectra for the distributions considered in Fig.~\ref{fig:Supra_All} (a,b) for $\langle |a_i| \rangle = 1$, ${\mu_T} =9$mV and $\sigma^2 = 4$ mV$^2$. In both panels, the symbols correspond to the simulation results, 
while the continuous line to the theoretical estimations using Eq. \eqref{eq:Cw}. Simulated spectra 
were obtained by calculating the squared modulus of the Fourier transform of the spike train using a 
time trace of 64 s with 1 ms binning, and averaging over 10,000 realizations.}
\label{fig:spectra}
\end{figure}

A much more detailed characterization of the firing time statistics,
beyond the first two moments that we have considered so far,
can be achieved by evaluating the spike train spectrum ${\hat C}(\omega)$ 
 which is directly related with the first-passage time density
(\ref{eq:Cw}). The comparison of the theoretical estimations with the numerical
findings is reported in Fig.~\ref{fig:spectra}, showing a very good agreement 
for all the reported cases. 
We report each spectra normalized by the corresponding average firing rate $r_0$, 
in order to emphasize the changes produced by the shape of the IPSP distribution,
rather than the changes due to the different values of the firing rates.
From Fig.~\ref{fig:spectra} (a), we observe that for $\delta$-distributed synaptic weights,
the increase in the kick amplitude $a_i$ induces a higher peak in the spectrum at a lower frequency.
Therefore, for increasing $a_i$ not only the rate
decreases, as previously reported, but also the 
peak of the ISI distribution shifts from 41 msec for $|a_i|=0.1$ mV to 54 msec
for $|a_i|=1$ mV, thus suggesting that the entire dynamics slows down
for increasing IPSP amplitudes.

Furthermore, as shown in Fig.~\ref{fig:spectra} (b)
the shape of the distributions of the IPSPs has also
an influence on the spectra. In particular, 
the increase in the asymmetry of the distributions induces
peaks at lower and lower frequencies, while their height increases. The 
neuron activity is slowed down for longer and longer tails in the
IPSP distributions.

\begin{figure}
\centering
\includegraphics[width=1\linewidth]{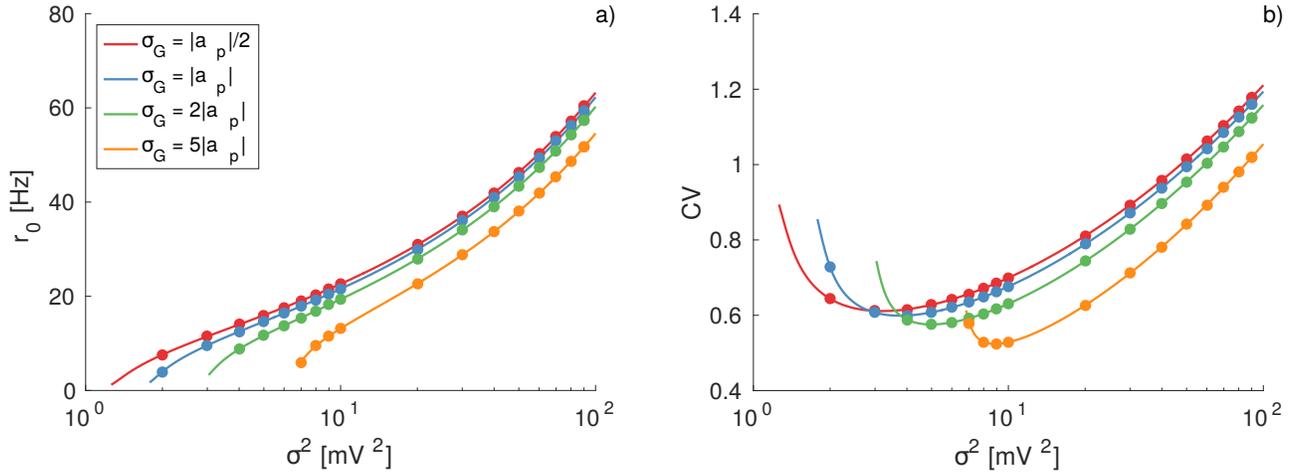}
\caption{{\bf Effect of the distribution asymmetry on the firing time statistics.} 
a) Average firing rate $r_0$ as a function of the noise intensity for different 
values of the standard deviation of the Gaussian distribution $\sigma_G$ as indicated in the legend. 
b) Coefficient of variation CV for the same cases depicted in a). The reported
data refer to TGDs with the peak located at $|a_p| = 1$ mV and to an effective input $\mu_T = 9$ mV.
Symbols correspond to numerical simulations, and the solid lines to the theoretical results reported in Eq.~\eqref{eq:firing_rate_generic} and Eq.~\eqref{eq:cv}. Other parameters and simulation procedures as in Fig. 
\ref{fig:Diffusion}}
\label{fig:asymmetry}
\end{figure}

These conclusions are supported by the data
reported in Fig.~\ref{fig:asymmetry} (a).
In the figure are reported the average firing rate
$r_0$ for TGDs with different standard deviation $\sigma_G$ and fixed position of the 
maximum at $a_p= -1$ mV. For increasing $\sigma_G$ the neuronal activity decreases
for corresponding noise amplitudes. Since the value of the skewness increases with
$\sigma_G$, more asymmetric IPSP distributions
induce a lower neuronal firing rate. 
Furthermore, a larger asymmetry is also responsible for
a shift of the position of the minimum of CV,
associated to the CR phenomenon, towards larger $\sigma^2$ and 
for a more regular firing activity observable at $\sigma^2 > \sigma_m^2$, 
as shown in  Fig.~\ref{fig:asymmetry} (b).

\begin{figure}
\centering
\includegraphics[width=1\linewidth]{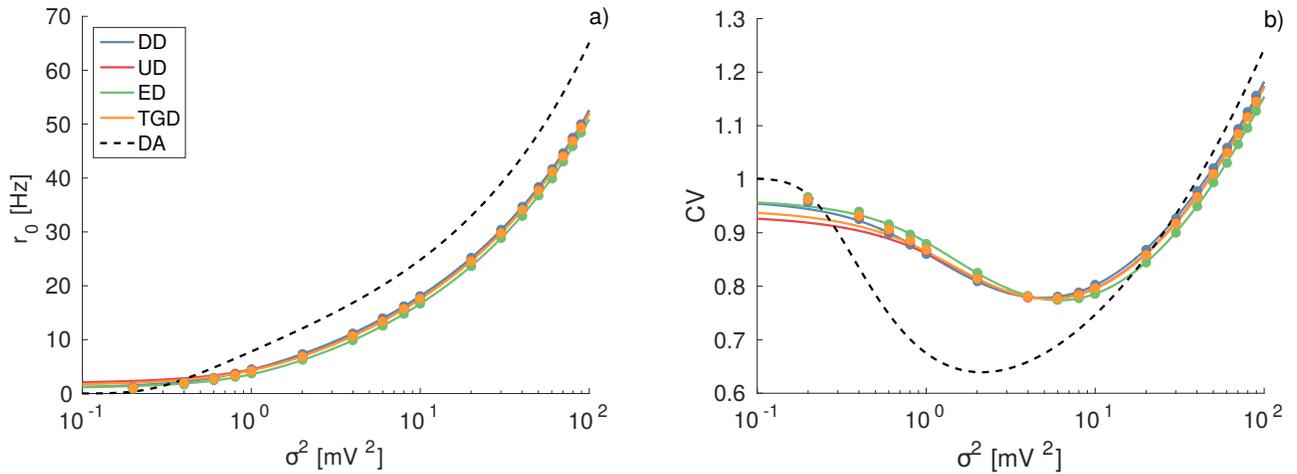}
\caption{{\bf Firing time statistics for exponentially distributed EPSPs.}
Firing rate $r_0$ (a) and  CV (b) as a function of noise intensity for excitatory synaptic weights distributed exponentially with average amplitude $\langle a_e \rangle = \langle | a_i | \rangle = 1$ mV. In this figure excitatory and inhibitory 
spike trains are balanced, i.e $R_e = R_i$, while the effective input is sub-threshold, namely $\mu_T = \mu_0 = 9$ mV. 
The symbols and lines have the same definition as in Fig. \ref{fig:Supra_All}, as well as all the other parameters for the IPSP distributions and the numerical simulations.}
\label{fig:Sub_All}
\end{figure}

 So far we considered as excitatory input a constant DC term, however
 the analytic approach here presented can be applied also for exponentially distributed
 excitatory amplitudes, namely $A_e = \exp(-a/a_e)/a_e$.
 We have reported the results for this case for different 
 inhibitory distributions in Fig.~\ref{fig:Sub_All}, the analytic
 estimations reproduce very well the numerical results 
 both for the firing rate and the CV in an ample range of noise intensities.
 However, for small values of the noise intensity, the analytic results slightly deviates from the
 numerical values due to the fact that in the limit of small rates
 the numerical evaluation of the integrals entering in the expresssion of the 
 generating function $Z_0$ have problems of convergece. The effect due to the different IPSP distributions 
 is analogous to the one observed with a  constant DC excitatory term. The DA, 
 conversely, is unable to capture the precise values of these two quantities. Once more, for the reported 
 case in Fig.~\ref{fig:Sub_All} where ${\mu_T} = 9$ mV, the CR effect is present.
 It is important to remark that, in the specific case of exponentially distributed amplitudes of the 
 EPSP, the approach discussed in this article fails when an additional supra-threshold DC current is applied, 
 namely for ${\mu_0} > v_{th}$. A brief discussion in this regard will be provided in the concluding section.

\subsection*{Heterogeneous Sparse Networks}

The previous theoretical analysis of single neuron response to an external Poissonian
input can find application also in the analysis of the dynamics of recurrent LIF heterogeneous networks
with random sparse connectivity. For sparse networks, the spike-trains
impinging a certain neuron can be
assumed to be uncorrelated and Poissonian~\cite{BrunelHakim1999,
Brunel2000Sparse}. Furthermore, similarly to what done in ~\cite{BrunelHakim1999},
we can assume that the spike-trains in the networks can be self-consistently
described as Poissonian processes with firing rates $r_0(\mu(j))$ related
to the neuronal excitability $\mu(j)$ of each single neuron.
 
For the sake of simplicity, we will consider a network of inhibitory neurons,
where each neuron is characterized by a different level of excitability $\{\mu(j)\}$ 
encompassing any excitatory external drive as well as the specific characteristic
of the considered neuron. Therefore, the dynamics of the $j$-th neuron in the network can be written as 
\begin{equation}
\label{eq:generic_model}
\dot{v}(j) = -\frac{[v(j) - \mu(j)]}{\tau} + \frac{1}{K}\sum_{k} C_{jk} g_i(k) \delta(t-t_k) \quad;
\end{equation}
where $K << N$ is the number of synaptic neighbors, $C_{jk}$ is the connectivity matrix with entries 1 (0) 
if the $k$-th neuron is connected (not connected) to neuron $j$. The amplitudes of the IPSP $a_i(k)= g_i(k)/K < 0$ 
associated to the firing of the $k$-th neuron is assumed to be
randomly distributed following some of the distributions previously introduced.

For the population dynamics we will give an estimate of the average firing rate via a self-consistent
approach by assuming that in average each neuron receive a single Poissonian spike train with a rate 
$R_i = \bar{r}_0 K$ and where each IPSP has a random amplitude $a_i$ taken from a distribution $A_i(a)$.
An estimation of the average firing rate can be obtained by solving the following implicit equation
\begin{equation}
\label{eq:nu_sparse}
\tau \bar{r}_0 =  \frac{1}{\Delta} \int_{\{\mu_A\}} d\mu P(\mu) \left[ \int_0^{\bar{x}} \frac{dc}{c}\frac{F(c)}{c Z_0(c,\mu,\bar{r}_0)} \right]^{-1}
\quad ;
\end{equation}
which is an extension to an heterogeneous network of Eq.~\eqref{eq:firing_rate_generic}
and where, as we have previously discussed, $F(c)$ depends on the excitatory input and $Z_0$ on the chosen IPSP distribution.
It is important to stress that usually in inhibitory networks not all neurons are firing, but just a certain
fraction $n^*$ will be active~\cite{olmi2016}, therefore the integral reported in \eqref{eq:nu_sparse} is limited
to these neurons, which are the one with higher excitability $\mu(j)\in\{\mu_A\}$ .
Furthermore $\Delta = \int_{\{\mu_A\}} d\mu P(\mu)$ is the support of the active neurons.
Once the firing rates have been obtained self consistently we can derive the coefficient of 
variation of each single neuron according to Eqs.~\eqref{eq:moments} and \eqref{eq:cv} and then 
to perform the population average as follows
\begin{equation}
\label{eq:cv_sparse}
\overline{\text{CV}} = \frac{1}{\Delta} \int_{\{\mu_A\}} d\mu P(\mu) \text{CV}(\mu)
\quad .
\end{equation}

\begin{figure}
\centering
\includegraphics[width=1\linewidth]{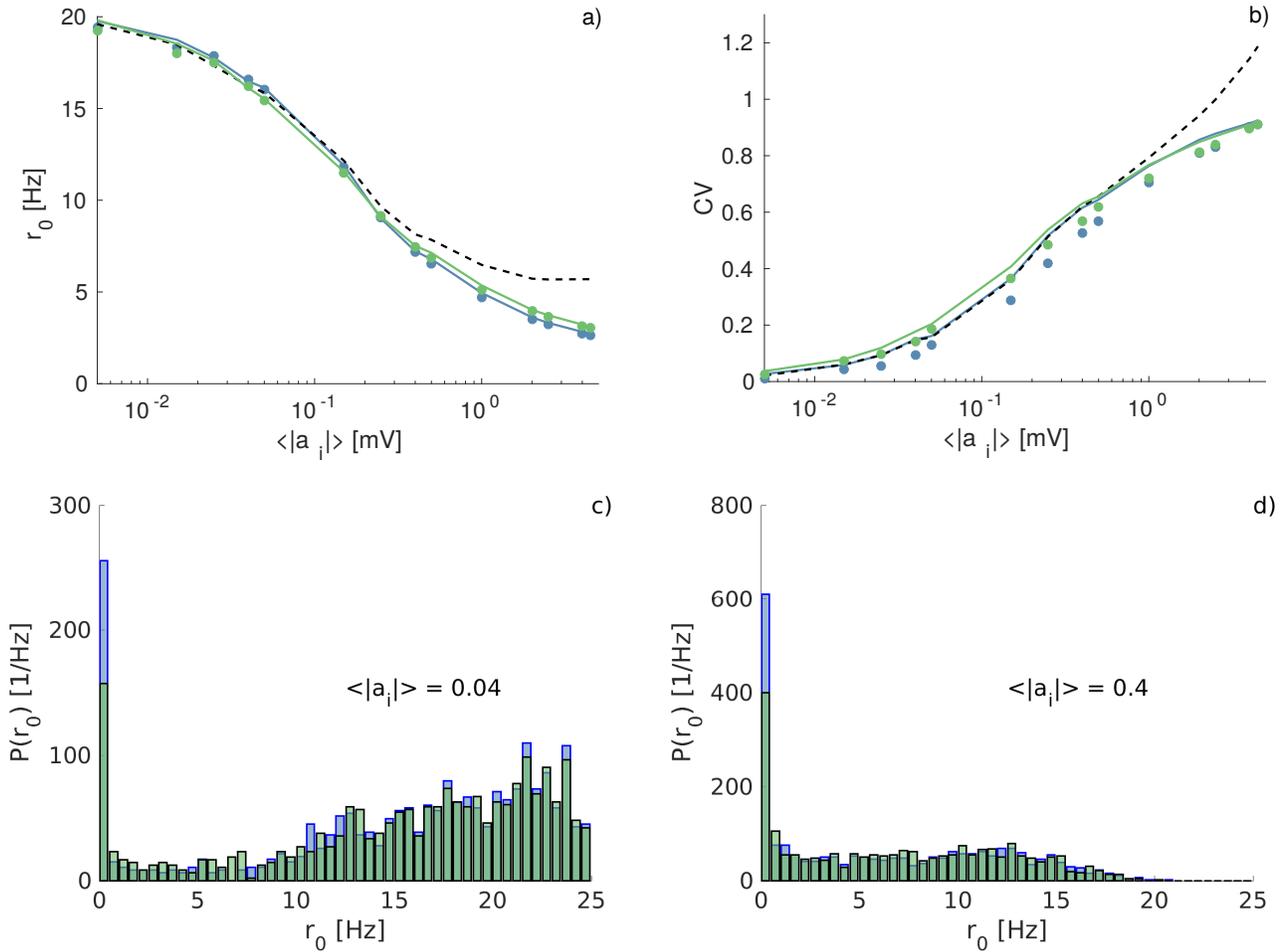}
\caption{{\bf Firing time statistics for heterogenoeus networks as a function of the average synaptic weight.} 
(a) Average network frequency $\bar r_0$ and (b)
average coefficient of variation $\overline{\text{CV}}$ as a function of the average synaptic weight
$\langle |a_i| \rangle$ for two representative IPSP distributions: DD (blue) and ED (green). 
Symbols correspond to simulation values and solid lines to the theoretical results from Eqs. 
\eqref{eq:nu_sparse}  and \eqref{eq:cv_sparse}. Dashed line denote the results of the DA. 
The numerically estimated probability distribution functions $P(r_0)$ of the single neuron 
firing rate $r_0$ are also reported for the two considered IPSP distributions
for two values of the average synaptic weight:
namely $\langle |a_i| \rangle=0.04$ (c) and $\langle |a_i| \rangle=0.4$ (d).
Simulations were performed with $N = 400$ neurons, 
$K = 20$ and a distribution of the input currents uniformly chosen in the interval $\mu(j) \in [10,\; 11]$ mV. 
The time averages are calculated, after discarding an initial transient corresponding to $10^6$ spikes, over the following $10^6$ spikes. Silent neurons (those that do not emit any spike in the considered time lapse) are not included in the statistics. 
In all cases $\langle |a_i| \rangle$ denotes the average value of the corresponding distribution.}
\label{population}
\end{figure}

In ~\cite{olmi2016} a theoretical approach to obtain self-consistently $n^*$, $\bar{r}_0$ and 
the average coefficient of variation $\overline{\text{CV}}$ has been developed for constant IPSP, 
i.e. for $\delta$-distribution.  Here we extend such approach to more generic IPSP distributions, 
however we will limit to obtain the analytic estimations of 
$\bar{r}_0$ and $\overline{\text{CV}}$.
The values of $n^*$, entering in the expressions for the average
population rate and coefficient of variation, will be considered as parameter values obtained directly from the simulations.
We have verified that the comparison with the numerical findings is very good for all the four considered 
IPSP distributions and for the whole considered range of average synaptic inputs $\langle a_i \rangle$. For
illustration purposes we present in Fig.~\ref{population} (a) and (b) the average firing rate $\bar{r}_0$ and 
$\overline{\text{CV}}$, for the two 
distributions that have consistently presented larger differences between them, namely DD and ED,
and for the DA. While it is evident that the DA overestimates $\bar{r}_0$ and $\overline{\text{CV}}$ already for
$\langle |a_i| \rangle > 0.5$ mV, we observe that in the case of heterogeneous networks the differences among the 
various IPSP distributions are quite limited  at the level of the  average firing rate 
$\bar{r}_0$ and $\overline{\text{CV}}$. In order to investigate
more in details the influence of different IPSP distributions on the neuronal
response,  we have numerically estimated the probability distribution functions $P(r_0)$ of the single neuron 
firing rate $r_0$ for DD and ED distributions. 
As shown in Fig.~\ref{population} (c) and (d), the $P(r_0)$ obtained
for the DD display a higher peak at low firing rate $r_0$ with respect to the ED,
while for higher firing rates they essentially coincide.
This difference should be ascribed to the fact that 
sub-threshold neurons subject to IPSP trains with ED have 
a firing onset at definitely larger noise amplitudes than the ones
subject to IPSPs with DD, while for supra-threshold neurons the differences
are definitely less evident, as previously reported in Fig.~\ref{fig:Supra_All} (a) and (c).
This explains also why the exam of the average firing rate does not reveal large
difference between the two distributions, since the neurons with
low firing rate contribute negligibly to the average activity.
The influence of synaptic weight distributions on neuronal 
population dynamics has been usually examined in the context of homogeneous neuronal populations~\cite{richardson2010firing,iyer2013}, where the single neuron response represents
a good mean field approximation of the network dynamics. However, from the present analysis
it emerges that the heterogeneity in the single neuron excitability can render extremely difficult to
distinguish among stimulations with different IPSP distributions.
  
As a final remark, we would like to stress that the reported approach works very well also
for heterogeneous sparse networks, provided that the collective dynamics is asynchronous.
Otherwise, the presence of partial synchronization or of collective oscillations
can induce correlations in the input spike trains, which cannot be accounted for
with this approach. In particular, to avoid phase locking among the neurons 
the distribution $P(\mu)$ of the excitabilities should be sufficiently wide.

\section*{Conclusions}

In this paper we have reported a theoretical methodology to obtain exact firing statistics
for leaky integrate-and-fire neurons subject to discrete inhibitory noise,
accounting for Poissonian trains of uncorrelated post-synaptic potentials.
Our results represent an extension to generic synaptic weights distribution of the
approach developed in~\cite{richardson2010firing} for exponentially distributed post-synaptic potentials.
In particular, we report explicit results for the firing rate, the coefficient of variation
and the spike train spectrum.

The comparison with numerical simulations reveals a very good agreement for all the considered
distributions over all the reported ranges of shot noise amplitude. Moreover, the method is also 
able to reproduce the average activity of an heterogeneous inhibitory neural network with sparse connectivity, 
by making use of a self-consistent mean field formulation. Conversely, the diffusion 
approximation~\cite{ricciardi2013diffusion} (the most used theoretical approach), gives a reasonable estimate
of the firing time statistics for sufficiently small IPSP amplitudes, but the agreement 
rapidly degrades, and reveals large discrepancies already for amplitudes $> 0.5$ mV, 
corresponding to physiologically relevant values~\cite{miles1990,buhl1994,tepper2004gabaergic}.

As a general result we observe that the firing statistics of single neurons is strongly influenced 
by the shape of the IPSP distributions. Distributions with longer tails lead to smaller firing rate 
and in general to more regular spike trains (for sufficiently large inhibition). 
For heterogeneous networks the value of the average IPSP has a noticeable influence
on the firing activity of the neuronal population, while the shape of the synaptic weight distributions seem to have 
a really limited impact on the average properties of the network. However, differences induced by
the IPSP distributions are still observable at the level of single neuron.
 
\begin{figure}
\centering
\includegraphics[width=0.45\linewidth]{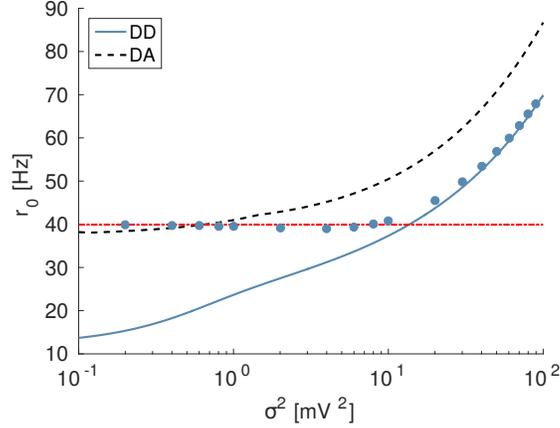}
\caption{{\bf Firing time statistics for exponentially distributed EPSPs and supra-threshold DC current}
Firing rate $r_0$ as a function of noise intensity for excitatory synaptic weights distributed exponentially and
a supra-threshold DC current. All parameters as in Fig. \ref{fig:Sub_All} except for $\mu_T = \mu_0 = 12$ mV.
The dashed red horizontal line refers to the rate $r_0^t$ of the LIF neuron subject to constant current 
$\mu_T$ (Eq. \eqref{eq:LIF_simple}). Other symbols and lines have the same meaning as in Fig. \ref{fig:Supra_All}, 
as well as all the other parameters for the IPSP distribution and the numerical simulations.} 
\label{fig:Drift_and_Expo}
\end{figure}

We have shown that the method works for any choice of IPSP amplitude distributions, 
however one has to be careful when dealing with the excitatory drive. We 
have shown that the agreement with numerical simulations is very good 
when the firing of the neuron is either promoted exclusively by a 
suprathreshold DC current or provided by the stochastic arrivals of exponentially 
distributed EPSP amplitudes. In this latter case, an excitatory DC current can as well
be present in addition to the EPSP trains, however the DC contribution must
be strictly sub-threshold, namely, $\mu_0 < v_{th}$. Whenever the two firing mechanisms are 
active at the same time (i.e. $\mu_0 > v_{th}$), 
the formulation reported in this paper is no more valid due to the inconsistency in the choice of the proper 
integration limit $\bar{x}$ in Eq.~\eqref{eq:firing_rate_generic}. This limitation is illustrated in 
Fig. \ref{fig:Drift_and_Expo}, where it is reported the response of a LIF neuron, subject to a supra-threshold 
excitatory DC current $\mu_0 =12$ mV as well as to Poissonian spike trains of ED excitatory and 
DD inhibitory synaptic inputs.
For small values of the noise intensity, the theoretical approach dramatically fails. 
In such a region the activity is almost exclusively 
current driven; i.e, the firing of the neuron promoted by the supra-threshold DC
current is much faster than the arrival rate $R_e$ of the EPSPs. This can be confirmed by the fact that in this  
regime the firing rate coincides with $r_0^t$ in Eq.~\eqref{eq:LIF_simple} for a
tonic firing LIF neuron subject to a constant DC current  $\mu_T=\mu_0$.
As the noise intensity grows, corresponding to an increased rate $R_e$ of arrival of the EPSPs, the numerical
data approach the theoretical prediction obtained for exponentially distributed EPSPs.
The crossover occurs for $R_e \approx 2  r_0^t$, indicating that the firing activity of the neuron
is now mainly driven by the stochastic component.

An interesting semi-analytic approach has been 
recently reported for excitatory shot noise in~\cite{iyer2013}.
However, further progresses are required to achieve exact firing time 
statistics going beyond the diffusion approximation
for generic distributions of instantaneous excitatory PSPs,
as well as for more realistic neuronal models like the Exponential Integrate and Fire (EIF) neuron~\cite{fourcaud} able to reproduce quite accurately the dynamics of cortical neurons~\cite{badel}.
In particular, as shown in the Supplementary Information, the LIF neuron response
is a limit case of the EIF dynamics both within the diffusion approximation
as well as for shot noise. Furthermore, the diffusion approximation fails also for the EIF to capture the neuronal firing statistics for sufficiently large IPSP amplitudes in particular at low noise intensities.

\section*{Acknowledgments}
We thank for useful discussions B. Lindner, G. Mato, A. Politi, G. Martelloni, MJE Richardson,  M. Timme. 
This work has been partially supported by the European Commission under the program ``Marie Curie Network for 
Initial Training", through the project N. 289146, ``Neural Engineering Transformative Technologies (NETT)"  
and by the A$^\ast$MIDEX grant (No. ANR-11-IDEX-0001-02) funded by the French Government ``Programme Investissements 
d'Avenir'' (D.A.-G. and A.T.). AT gratefully acknowledges a ``VELUX Visiting Professor Grant" from Villum 
Fonden that supported his stay at Aarhus University during the initial stages of this work in 2011. AI 
is supported by the  Danish Council for Independent Research and the Villum Fonden. This work has been 
completed at Max Planck Institute for the Physics of Complex Systems in Dresden (Germany) as part of the 
activity of the Advanced Study Group 2016 entitled "From Microscopic to Collective Dynamics in Neural Circuits".

\section*{Author contributions statement}

S.O and D.A-G performed the numerical simulations. D.A-G and A.T prepared the manuscript. All the authors developed
the theoretical methods and reviewed the manuscript.

\section*{Additional information}

\textbf{Competing financial interests} Authors declare to have no competing financial
interests that might have influenced this manuscript.

\section*{Supplementary Information}

\section*{Exponential Integrate and Fire}

The Exponential Integrate-and-Fire (EIF)  model is a simple non-linear integrate and fire neuronal
model introduced by Fourcaud-Trocm\'e et al.~\cite{fourcaud} able to reproduce
quite accurately the dynamics of cortical neurons~\cite{badel}.
The model can be written as follows
\begin{equation}
\label{eq:EIF}
\tau \dfrac{dv}{dt} = \mu_0 - v + \Delta_T \exp\left(\frac{v - v_{th}}{\Delta_T}\right) + I(t) \;.
\end{equation}
where $\mu_0$ represents an external DC current and $I(t)$ the synaptic drive. 
In this model, the spike generation occurs in a finite time controlled by the parameter $\Delta_T$.
In particular, once the membrane potential has reached the threshold value $v_{th}$ this will rapidly 
grow towards infinity in a finite time interval, the parameter $\Delta_T$ establishes how fast the infinite limit is 
reached. In the limit ${\Delta_T \rightarrow 0}$, the spike generation is instantaneous and the LIF model is recovered.
As in the usual LIF model, once the neuron has fired its membrane potential is resetted to the value
$v_{re} = 5$ mV, we also set $\tau = 20$ ms and $v_{th} = 10$ mV, as in the LIF model studied in the 
article.

As a first analysis, we will examine the response of the EIF neuron subject to small Gaussian noise of zero
average and intensity $\sigma$, this can be obtained by solving the associated 
continuity equation for the probability $P(v,t)$ of finding the membrane voltage between $v$ and $v+dv$
at time $t$, which reads as~\cite{fourcaud}:
\begin{eqnarray}
\label{eq:fokker}
\frac{\partial P}{\partial t} + \frac{\partial J}{\partial v}& = & r(t)[\delta(v-v_{re}) - \delta(v-v_{th})] - \delta(t)\delta(v-v_{re}) \\
\label{eq:flux}
J & = & \left( \dfrac{\mu_0 -v + \Delta_T {\rm e}^{v-v_{th}/\Delta_T}}{\tau} \right)P
- \dfrac{\sigma^2}{2\tau} \frac{\partial P}{\partial v} 
\end{eqnarray}
where $J=J(v,t)$ is the associated flux. .
In this case, since the effective threshold is located at infinity, the steady firing
rate can be evaluated as 
$$r_0 = \lim_{v \to \infty} J(v) \; ;$$
where $J(v)$ is the stationary solution of the continuity equation for the flux.

\begin{figure}
\centering
\includegraphics[width=0.45\linewidth]{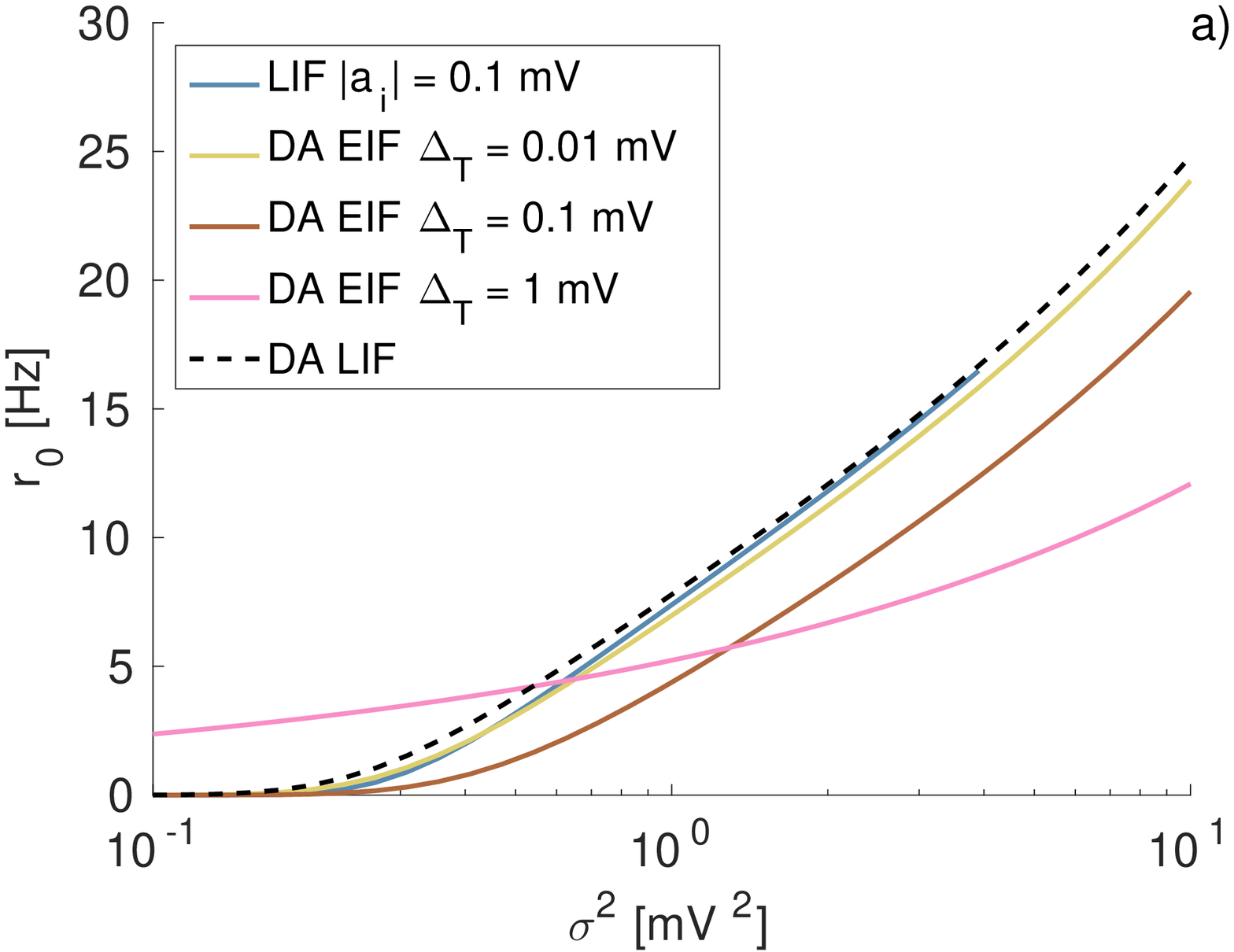}
\includegraphics[width=0.45\linewidth]{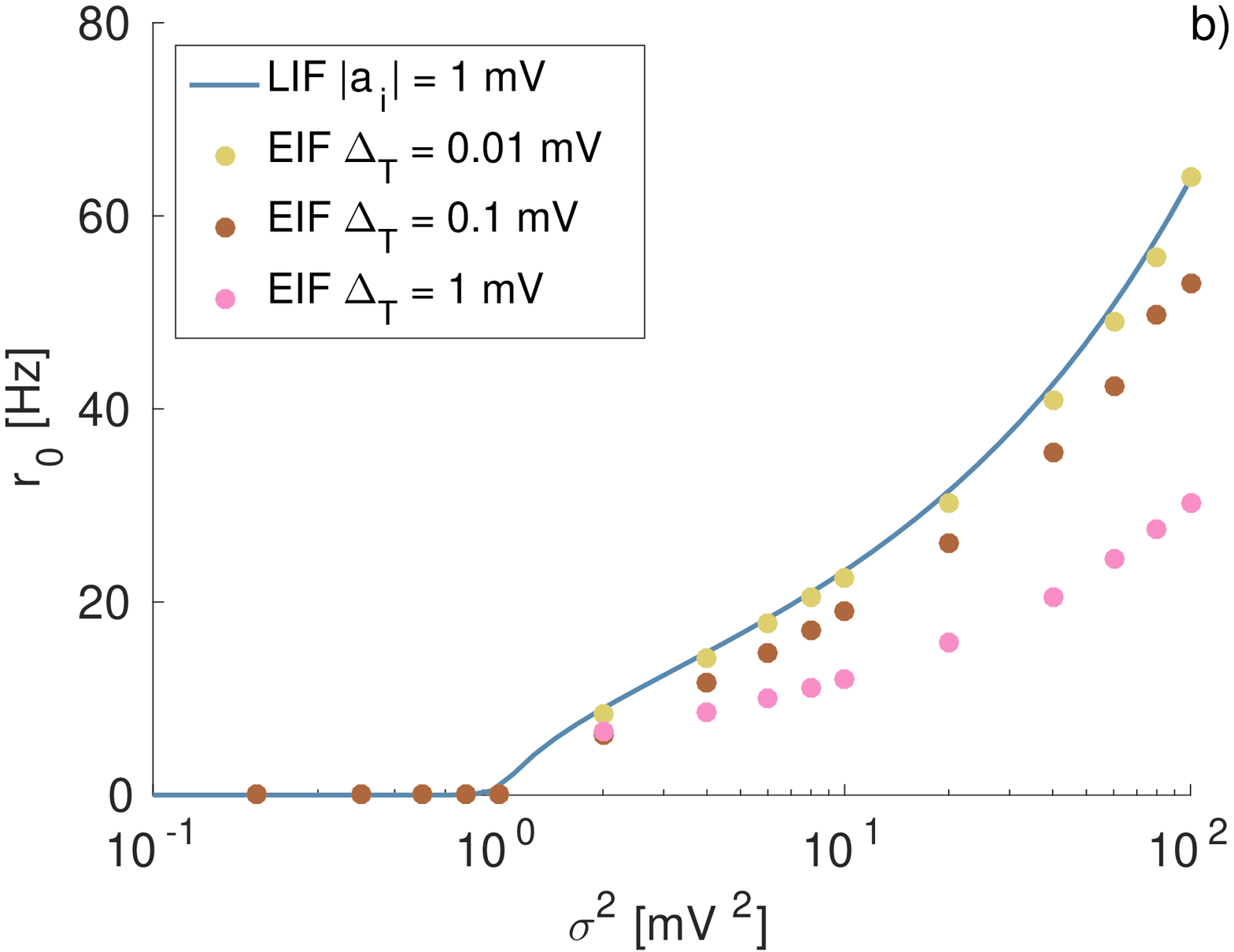}
\caption{\textbf{Comparison of the firing rate between LIF and EIF} 
a) Firing rate in the Diffusion limit for the EIF at different $\Delta_T$ and the
comparison with the DA and $\delta$-distributed amplitudes of IPSP in the LIF model with small $|a_i| = 0.1$ mV. 
Results for the EIF are obtained by solving the stationary state of Eqs. \eqref{eq:fokker} and \eqref{eq:flux} via the
threshold integration method as reported in \cite{richardson2007}. For the LIF, the DA and the $\delta$-distributed 
solutions are taken respectively from Eqs. (17) and (51) in the main text.
b) Numerical simulations of the EIF in the shot noise case with $\delta$-distributed IPSP amplitudes 
of average $|a_i| = 1$mV, for the same values of $\Delta_T$ as in panel a), and the corresponding LIF case with
the same IPSP distribution. In this panel, the  results of the EIF are calculated numerically by 
integrating Eq. \eqref{eq:EIF} with an Euler scheme 
with time step $h = 1\times 10^{-3}$. When the neuron reaches a large value $v_{\infty} = 80$ mV, the remaining 
time to reach infinity is calculated as $t_{\infty} = \tau \exp((v_{th}-v_{\infty})/\Delta_T)$ \cite{fourcaud}. 
In all the cases we have chosen $\mu_T = 9$mV, $v_{re} = 5$mV, $v_{th} = 10$mV and $\tau = 20$ms.}
\label{fig:Diffusion_LIF_vs_EIF}
\end{figure}

In particular, we made use of the threshold-integration method~\cite{richardson2007} to calculate the firing rate
$r_0$ of the EIF neuron subject to inhibitory inputs 
and compare it with the diffusion approximation (DA) for the LIF and the corresponding shot noise solution 
for $\delta$-distributed IPSP amplitudes with $|a_i| = 0.1$ mV. The results are shown in
Fig.~\ref{fig:Diffusion_LIF_vs_EIF} (a), where it is clearly shown that for $\Delta_T \to 0$ 
the EIF results converge to the LIF solution, both for the DA and the the shot noise results found for 
small IPSP. Furthermore direct simulations of the EIF and the corresponding shot noise solution of the 
LIF for large IPSP amplitudes, namely $|a_i| = 1$ mV, show that also in this case the LIF limit is 
recovered for $\Delta_T \to 0$. However it is clear that also in the case of the EIF, the diffusion limit
is unable to capture the onset of the activity and the firing rate at small intensities.

\begin{thebibliography}{10}
\expandafter\ifx\csname url\endcsname\relax
  \def\url#1{\texttt{#1}}\fi
\expandafter\ifx\csname urlprefix\endcsname\relax\def\urlprefix{URL }\fi
\expandafter\ifx\csname doiprefix\endcsname\relax\def\doiprefix{DOI }\fi
\providecommand{\bibinfo}[2]{#2}
\providecommand{\eprint}[2][]{\url{#2}}

\bibitem{destexhe2003}
\bibinfo{author}{Destexhe, A.}, \bibinfo{author}{Rudolph, M.} \&
  \bibinfo{author}{Par{\'e}, D.}
\newblock \bibinfo{title}{The high-conductance state of neocortical neurons in
  vivo}.
\newblock \emph{\bibinfo{journal}{Nature reviews neuroscience}}
  \textbf{\bibinfo{volume}{4}}, \bibinfo{pages}{739--751}
  (\bibinfo{year}{2003}).

\bibitem{shadlen1998}
\bibinfo{author}{Shadlen, M.~N.} \& \bibinfo{author}{Newsome, W.~T.}
\newblock \bibinfo{title}{The variable discharge of cortical neurons:
  implications for connectivity, computation, and information coding}.
\newblock \emph{\bibinfo{journal}{The Journal of neuroscience}}
  \textbf{\bibinfo{volume}{18}}, \bibinfo{pages}{3870--3896}
  (\bibinfo{year}{1998}).

\bibitem{ricciardi2013diffusion}
\bibinfo{author}{Ricciardi, L.~M.}
\newblock \emph{\bibinfo{title}{Diffusion processes and related topics in
  biology}}, vol.~\bibinfo{volume}{14} (\bibinfo{publisher}{Springer Science \&
  Business Media}, \bibinfo{year}{2013}).

\bibitem{tuckwell2005}
\bibinfo{author}{Tuckwell, H.~C.}
\newblock \emph{\bibinfo{title}{Introduction to theoretical neurobiology:
  Volume 2, nonlinear and stochastic theories}}, vol.~\bibinfo{volume}{8}
  (\bibinfo{publisher}{Cambridge University Press}, \bibinfo{year}{2005}).

\bibitem{roxin2011}
\bibinfo{author}{Roxin, A.}, \bibinfo{author}{Brunel, N.},
  \bibinfo{author}{Hansel, D.}, \bibinfo{author}{Mongillo, G.} \&
  \bibinfo{author}{van Vreeswijk, C.}
\newblock \bibinfo{title}{On the distribution of firing rates in networks of
  cortical neurons}.
\newblock \emph{\bibinfo{journal}{The Journal of neuroscience}}
  \textbf{\bibinfo{volume}{31}}, \bibinfo{pages}{16217--16226}
  (\bibinfo{year}{2011}).

\bibitem{song2005}
\bibinfo{author}{Song, S.}, \bibinfo{author}{Sj{\"o}str{\"o}m, P.~J.},
  \bibinfo{author}{Reigl, M.}, \bibinfo{author}{Nelson, S.} \&
  \bibinfo{author}{Chklovskii, D.~B.}
\newblock \bibinfo{title}{Highly nonrandom features of synaptic connectivity in
  local cortical circuits}.
\newblock \emph{\bibinfo{journal}{PLoS Biol}} \textbf{\bibinfo{volume}{3}},
  \bibinfo{pages}{e68} (\bibinfo{year}{2005}).

\bibitem{lefort2009}
\bibinfo{author}{Lefort, S.}, \bibinfo{author}{Tomm, C.},
  \bibinfo{author}{Sarria, J.-C.~F.} \& \bibinfo{author}{Petersen, C.~C.}
\newblock \bibinfo{title}{The excitatory neuronal network of the c2 barrel
  column in mouse primary somatosensory cortex}.
\newblock \emph{\bibinfo{journal}{Neuron}} \textbf{\bibinfo{volume}{61}},
  \bibinfo{pages}{301--316} (\bibinfo{year}{2009}).

\bibitem{miles1990}
\bibinfo{author}{Miles, R.}
\newblock \bibinfo{title}{Variation in strength of inhibitory synapses in the
  ca3 region of guinea-pig hippocampus in vitro.}
\newblock \emph{\bibinfo{journal}{The Journal of Physiology}}
  \textbf{\bibinfo{volume}{431}}, \bibinfo{pages}{659} (\bibinfo{year}{1990}).

\bibitem{mason1991}
\bibinfo{author}{Mason, A.}, \bibinfo{author}{Nicoll, A.} \&
  \bibinfo{author}{Stratford, K.}
\newblock \bibinfo{title}{Synaptic transmission between individual pyramidal
  neurons of the rat visual cortex in vitro}.
\newblock \emph{\bibinfo{journal}{The Journal of Neuroscience}}
  \textbf{\bibinfo{volume}{11}}, \bibinfo{pages}{72--84}
  (\bibinfo{year}{1991}).

\bibitem{barbour2007}
\bibinfo{author}{Barbour, B.}, \bibinfo{author}{Brunel, N.},
  \bibinfo{author}{Hakim, V.} \& \bibinfo{author}{Nadal, J.-P.}
\newblock \bibinfo{title}{What can we learn from synaptic weight
  distributions?}
\newblock \emph{\bibinfo{journal}{TRENDS in Neurosciences}}
  \textbf{\bibinfo{volume}{30}}, \bibinfo{pages}{622--629}
  (\bibinfo{year}{2007}).
 
\bibitem{deweese2006}
M. R. DeWeese and A. M. Zador, Non-Gaussian membrane potential dynamics imply sparse, synchronous activity in auditory cortex,
{\em Journal of Neuroscience} {\bf 26}, 12206-12218 (2006).
 
  
\bibitem{buzsaki2014}
\bibinfo{author}{Buzs{\'a}ki, G.} \& \bibinfo{author}{Mizuseki, K.}
\newblock \bibinfo{title}{The log-dynamic brain: how skewed distributions
  affect network operations}.
\newblock \emph{\bibinfo{journal}{Nature Reviews Neuroscience}}
  \textbf{\bibinfo{volume}{15}}, \bibinfo{pages}{264--278}
  (\bibinfo{year}{2014}).

\bibitem{iyer2013}
\bibinfo{author}{Iyer, R.}, \bibinfo{author}{Menon, V.},
  \bibinfo{author}{Buice, M.}, \bibinfo{author}{Koch, C.} \&
  \bibinfo{author}{Mihalas, S.}
\newblock \bibinfo{title}{The influence of synaptic weight distribution on
  neuronal population dynamics}.
\newblock \emph{\bibinfo{journal}{PLoS Comput Biol}}
  \textbf{\bibinfo{volume}{9}}, \bibinfo{pages}{e1003248}
  (\bibinfo{year}{2013}).

\bibitem{teramae2012}
\bibinfo{author}{Teramae, J.-n.}, \bibinfo{author}{Tsubo, Y.} \&
  \bibinfo{author}{Fukai, T.}
\newblock \bibinfo{title}{Optimal spike-based communication in excitable
  networks with strong-sparse and weak-dense links}.
\newblock \emph{\bibinfo{journal}{Scientific Reports}}
  \textbf{\bibinfo{volume}{2}} (\bibinfo{year}{2012}).

\bibitem{collins1995}
\bibinfo{author}{Collins, J.}, \bibinfo{author}{Chow, C.~C.},
  \bibinfo{author}{Imhoff, T.~T.} \emph{et~al.}
\newblock \bibinfo{title}{Stochastic resonance without tuning}.
\newblock \emph{\bibinfo{journal}{Nature}} \textbf{\bibinfo{volume}{376}},
  \bibinfo{pages}{236--238} (\bibinfo{year}{1995}).

\bibitem{gammaitoni1998}
\bibinfo{author}{Gammaitoni, L.}, \bibinfo{author}{H{\"a}nggi, P.},
  \bibinfo{author}{Jung, P.} \& \bibinfo{author}{Marchesoni, F.}
\newblock \bibinfo{title}{Stochastic resonance}.
\newblock \emph{\bibinfo{journal}{Reviews of modern physics}}
  \textbf{\bibinfo{volume}{70}}, \bibinfo{pages}{223} (\bibinfo{year}{1998}).

\bibitem{richardson2010firing}
\bibinfo{author}{Richardson, M.~J.} \& \bibinfo{author}{Swarbrick, R.}
\newblock \bibinfo{title}{Firing-rate response of a neuron receiving excitatory
  and inhibitory synaptic shot noise}.
\newblock \emph{\bibinfo{journal}{Physical review letters}}
  \textbf{\bibinfo{volume}{105}}, \bibinfo{pages}{178102}
  (\bibinfo{year}{2010}).

\bibitem{burkitt2006LIFreviewI}
\bibinfo{author}{Burkitt, A.~N.}
\newblock \bibinfo{title}{A review of the integrate-and-fire neuron model: I.
  homogeneous synaptic input}.
\newblock \emph{\bibinfo{journal}{Biological cybernetics}}
  \textbf{\bibinfo{volume}{95}}, \bibinfo{pages}{1--19} (\bibinfo{year}{2006}).

\bibitem{burkitt2006LIFreviewII}
\bibinfo{author}{Burkitt, A.~N.}
\newblock \bibinfo{title}{A review of the integrate-and-fire neuron model: Ii.
  inhomogeneous synaptic input and network properties}.
\newblock \emph{\bibinfo{journal}{Biological cybernetics}}
  \textbf{\bibinfo{volume}{95}}, \bibinfo{pages}{97--112}
  (\bibinfo{year}{2006}).

\bibitem{stein1965}
\bibinfo{author}{Stein, R.~B.}
\newblock \bibinfo{title}{A theoretical analysis of neuronal variability}.
\newblock \emph{\bibinfo{journal}{Biophysical Journal}}
  \textbf{\bibinfo{volume}{5}}, \bibinfo{pages}{173} (\bibinfo{year}{1965}).
 
\bibitem{rauch} Rauch A., La Camera G., Luscher H-R., Senn W. and Fusi S.,
Neocortical Pyramidal Cells Respond as Integrate-and-Fire Neurons to In Vivo Like Input Currents, 
{\em J. Neurophys} {\bf 90} 1598-1612 (2003).

\bibitem{jolivet} Jolivet A., Rauch A, L\"uscher H-R., Gerstner W. 
Predicting spike timing of neocortical pyramidal neurons by
simple threshold models, {\em J Comput Neurosci} {\bf 21} 35-49. (2006).











\bibitem{gerstner2002spiking}
\bibinfo{author}{Gerstner, W.} \& \bibinfo{author}{Kistler, W.~M.}
\newblock \emph{\bibinfo{title}{Spiking neuron models: Single neurons,
  populations, plasticity}} (\bibinfo{publisher}{Cambridge university press},
  \bibinfo{year}{2002}).

\bibitem{richardson2008spike}
\bibinfo{author}{Richardson, M.~J.}
\newblock \bibinfo{title}{Spike-train spectra and network response functions
  for non-linear integrate-and-fire neurons}.
\newblock \emph{\bibinfo{journal}{Biological Cybernetics}}
  \textbf{\bibinfo{volume}{99}}, \bibinfo{pages}{381--392}
  (\bibinfo{year}{2008}).

\bibitem{tuckwell1976first}
\bibinfo{author}{Tuckwell, H.~C.}
\newblock \bibinfo{title}{On the first-exit time problem for temporally
  homogeneous markov processes}.
\newblock \emph{\bibinfo{journal}{Journal of Applied Probability}}
  \bibinfo{pages}{39--48} (\bibinfo{year}{1976}).

\bibitem{giraudo1997jump}
\bibinfo{author}{Giraudo, M.~T.} \& \bibinfo{author}{Sacerdote, L.}
\newblock \bibinfo{title}{Jump-diffusion processes as models for neuronal
  activity}.
\newblock \emph{\bibinfo{journal}{Biosystems}} \textbf{\bibinfo{volume}{40}},
  \bibinfo{pages}{75--82} (\bibinfo{year}{1997}).

\bibitem{kou2003first}
\bibinfo{author}{Kou, S.~G.} \& \bibinfo{author}{Wang, H.}
\newblock \bibinfo{title}{First passage times of a jump diffusion process}.
\newblock \emph{\bibinfo{journal}{Advances in applied probability}}
  \bibinfo{pages}{504--531} (\bibinfo{year}{2003}).

\bibitem{lindner2002}
\bibinfo{author}{Lindner, B.}, \bibinfo{author}{Schimansky-Geier, L.} \&
  \bibinfo{author}{Longtin, A.}
\newblock \bibinfo{title}{Maximizing spike train coherence or incoherence in
  the leaky integrate-and-fire model}.
\newblock \emph{\bibinfo{journal}{Physical Review E}}
  \textbf{\bibinfo{volume}{66}}, \bibinfo{pages}{031916}
  (\bibinfo{year}{2002}).

\bibitem{BrunelHakim1999}
\bibinfo{author}{Brunel, N.} \& \bibinfo{author}{Hakim, V.}
\newblock \bibinfo{title}{Fast global oscillations in networks of
  integrate-and-fire neurons with low firing rates}.
\newblock \emph{\bibinfo{journal}{Neural. Comput.}}
  \textbf{\bibinfo{volume}{11}}, \bibinfo{pages}{1621--1671}
  (\bibinfo{year}{1999}).

\bibitem{amit1997}
\bibinfo{author}{Amit, D.~J.} \& \bibinfo{author}{Brunel, N.}
\newblock \bibinfo{title}{Model of global spontaneous activity and local
  structured activity during delay periods in the cerebral cortex.}
\newblock \emph{\bibinfo{journal}{Cerebral cortex}}
  \textbf{\bibinfo{volume}{7}}, \bibinfo{pages}{237--252}
  (\bibinfo{year}{1997}).

\bibitem{dummer2014}
\bibinfo{author}{Dummer, B.}, \bibinfo{author}{Wieland, S.} \&
  \bibinfo{author}{Lindner, B.}
\newblock \bibinfo{title}{Self-consistent determination of the spike-train
  power spectrum in a neural network with sparse connectivity}.
\newblock \emph{\bibinfo{journal}{Frontiers in computational neuroscience}}
  \textbf{\bibinfo{volume}{8}} (\bibinfo{year}{2014}).

\bibitem{lindner2004}
\bibinfo{author}{Lindner, B.}, \bibinfo{author}{Garc{\i}a-Ojalvo, J.},
  \bibinfo{author}{Neiman, A.} \& \bibinfo{author}{Schimansky-Geier, L.}
\newblock \bibinfo{title}{Effects of noise in excitable systems}.
\newblock \emph{\bibinfo{journal}{Physics Reports}}
  \textbf{\bibinfo{volume}{392}}, \bibinfo{pages}{321--424}
  (\bibinfo{year}{2004}).

\bibitem{pikovsky1997}
\bibinfo{author}{Pikovsky, A.~S.} \& \bibinfo{author}{Kurths, J.}
\newblock \bibinfo{title}{Coherence resonance in a noise-driven excitable
  system}.
\newblock \emph{\bibinfo{journal}{Physical Review Letters}}
  \textbf{\bibinfo{volume}{78}}, \bibinfo{pages}{775} (\bibinfo{year}{1997}).

\bibitem{olmi2016}
\bibinfo{author}{Angulo-Garcia, D.}, \bibinfo{author}{Luccioli, S.},
  \bibinfo{author}{Olmi, S.} \& \bibinfo{author}{Torcini, A.}
\newblock \bibinfo{title}{Death and rebirth of neural activity in sparse
  inhibitory networks}.
\newblock \emph{\bibinfo{journal}{preprint bioRxiv 082974}}
  (\bibinfo{year}{2016}).

\bibitem{Brunel2000Sparse}
\bibinfo{author}{Brunel, N.}
\newblock \bibinfo{title}{Dynamics of sparsely connected networks of excitatory
  and inhibitory spiking neurons}.
\newblock \emph{\bibinfo{journal}{J. Comput. Neurosci.}}
  \textbf{\bibinfo{volume}{8}}, \bibinfo{pages}{183--208}
  (\bibinfo{year}{2000}).

\bibitem{buhl1994}
\bibinfo{author}{Buhl, E.~H.}, \bibinfo{author}{Halasy, K.} \&
  \bibinfo{author}{Somogyi, P.}
\newblock \bibinfo{title}{Diverse sources of hippocampal unitary inhibitory
  postsynaptic potentials and the number of synaptic release sites.}
\newblock \emph{\bibinfo{journal}{Nature}} \textbf{\bibinfo{volume}{368}},
  \bibinfo{pages}{823--828} (\bibinfo{year}{1994}).

\bibitem{tepper2004gabaergic}
\bibinfo{author}{Tepper, J.~M.}, \bibinfo{author}{Ko{\'o}s, T.} \&
  \bibinfo{author}{Wilson, C.~J.}
\newblock \bibinfo{title}{Gabaergic microcircuits in the neostriatum}.
\newblock \emph{\bibinfo{journal}{Trends in neurosciences}}
  \textbf{\bibinfo{volume}{27}}, \bibinfo{pages}{662--669}
  (\bibinfo{year}{2004}).

\bibitem{isope2002}
\bibinfo{author}{Isope, P.} \& \bibinfo{author}{Barbour, B.}
\newblock \bibinfo{title}{Properties of unitary granule cell --> purkinje cell
  synapses in adult rat cerebellar slices}.
\newblock \emph{\bibinfo{journal}{The Journal of neuroscience}}
  \textbf{\bibinfo{volume}{22}}, \bibinfo{pages}{9668--9678}
  (\bibinfo{year}{2002}).
  
\bibitem{fourcaud} Fourcaud-Trocm\'e N., Hansel D., van Vresswijk C., and Brunel N.,
How Spike Generation Mechanisms Determine the
Neuronal Response to Fluctuating Inputs, {\em J. Neurosci.} 
{\bf 23} 11628-11640 (2003).

\bibitem{badel} Badel L., Lefort S., Brette R., Petersen C.C.H., Gerstner W.,and Richardson M.J.E.,
Dynamic I-V curves are reliable predictors of naturalistic pyramidal-neuron voltage
traces, {\em J. Neurophysiol.} {\bf 99} 656-666 (2008).

  
  

\end{thebibliography}

\begin{thebibliography}{10}


\bibitem{fourcaud} Fourcaud-Trocm\'e N., Hansel D., van Vresswijk C., and Brunel N.,
How Spike Generation Mechanisms Determine the
Neuronal Response to Fluctuating Inputs, {\em J. Neurosci.} 
{\bf 23} 11628-11640 (2003).

\bibitem{badel} Badel L., Lefort S., Brette R., Petersen C.C.H., Gerstner W.,and Richardson M.J.E.,
Dynamic I-V curves are reliable predictors of naturalistic pyramidal-neuron voltage
traces, {\em J. Neurophysiol.} {\bf 99} 656-666 (2008).

\bibitem{richardson2007} Richardson M.J.E.,
Firing-rate response of linear and nonlinear integrate-and-fire neurons to modulated
current-based and conductance-based synaptic drive, {\em Phys. Rev. E 76}
021919 (2007).
 


\end{thebibliography}
\end{document}